\documentclass[%
superscriptaddress,
preprint,
 amsmath,amssymb,
 aps,
]{revtex4-2}

\usepackage{hyperref}
\usepackage{graphicx}% Include figure files
\usepackage{dcolumn}% Align table columns on decimal point
\usepackage{bm}% bold math
\usepackage{siunitx}
\usepackage{caption}
\usepackage[thinc]{esdiff}
\usepackage{comment}
\usepackage{verbatim}
\usepackage{float}
\usepackage{todonotes}

\newcommand*\chem[1]{\ensuremath{\mathrm{#1}}}
\newcommand{\Mean}[1]{\langle #1 \rangle}

\usepackage{subfig}
\usepackage{xr}[=v6]
\begin{document}

% \preprint{APS/123-QED}

\title{Compressibility of Confined Fluids from Volume Fluctuations}

\author{Jason Ogbebor}
\affiliation{Department of Materials Science and Engineering,\\
Massachusetts Institute of Technology,\\
77 Massachusetts Avenue, Cambridge, MA 02139, USA}
\author{Santiago A. Flores~Roman}
\affiliation{Otto H. York Department of Chemical and Materials Engineering,\\
New Jersey Institute of Technology,\\
323 Dr. Martin Luther King Jr. Blvd, Newark, NJ 07102, USA}
\author{Geordy Jomon}
\affiliation{Otto H. York Department of Chemical and Materials Engineering,\\
New Jersey Institute of Technology,\\
323 Dr. Martin Luther King Jr. Blvd, Newark, NJ 07102, USA}
\author{Gennady Y. Gor}
\email{E-mail: gor@njit.edu}
\affiliation{Otto H. York Department of Chemical and Materials Engineering,\\
New Jersey Institute of Technology,\\
323 Dr. Martin Luther King Jr. Blvd, Newark, NJ 07102, USA}

\begin{abstract}
When fluids are confined in nanopores, many of their properties deviate from bulk. These include bulk modulus, or compressibility, which determines the mechanical properties of fluid-saturated porous solids. Such properties are of importance for exploration and recovery of coal-bed methane and shale gas. We developed a new molecular simulation method for calculating compressibility of confined fluids, and applied it to methane in carbon nanopores. The method is based on volume fluctuations in the isothermal-isobaric ensemble, made possible through integrated potentials. Our method is one order of magnitude faster than the Monte Carlo approach, and allows calculations for pore sizes up to 100 nm. Our simulations predicted an increase in the fluid bulk modulus by a factor of 4 in 3 nm slit pores, and showed a gradual decrease with the increase of the pore size, so that at 100 nm, the deviation from the bulk is less than 5\%.
\end{abstract}

\maketitle

\pagebreak

\section{Introduction}

The phenomenon of changing fluid properties under nano-confinement has been studied by a number of works which have demonstrated significant alteration in density, phase transition temperatures, diffusivity, and other fluid properties~\cite{Huber2015, Gubbins2014, An2023}. The magnitude of these effects depends on the pore size and the strength of solid-fluid interactions~\cite{An2023}. Compressibility or bulk modulus is one of the properties altered by confinement~\cite{Dobrzanski2021}. Bulk moduli of both solid and fluid components determine the poroelastic behavior of fluid-saturated rocks, relevant to exploration and extraction of hydrocarbons~\cite{cheng2016poroelasticity}. 

Ultrasound propagation experiments on fluid-saturated nanoporous media have shown significant deviation in the bulk modulus of several fluids under confinement~\cite{Page1995, Schappert2013JoP, Schappert2013N2, Schappert2014, Schappert2014thesis, Ogbebor2023, Schappert2024, didier2025acoustic}. It has also been confirmed by grand canonical Monte Carlo (GCMC) simulations of argon, nitrogen, and methane in pores of various shapes. These studies showed that for the pores in the range of sizes 2-10 nm, the bulk modulus shows a linear dependence on the inverse of the pore diameter~\cite{Dobrzanski2018, Maximov2018, Corrente2020}. This mirrors the effect of confinement on the freezing point given by the Gibbs-Thomson equation~\cite{An2025, An2023}, but the observed trend did not intersect with the modulus of the bulk fluid at the limit of an infinitely large pore, suggesting that the true dependence should eventually plateau with increasing pore size. While calculations of bulk modulus of confined fluids using GCMC are straightforward, the computational cost becomes prohibitive for large pore sizes. This limitation motivates the development of alternative simulation methods to investigate the effects of confinement on the bulk modulus of fluids in a pore of ever increasing size. In this letter, we introduce a method for calculating the fluid bulk modulus using a well-established relation from statistical mechanics which previously could only be applied to bulk fluids, and for which the speed of the calculation scales well with the size of the pore.

%\section{Current Methods}

To put this new approach into context, it is appropriate to briefly take account of the current established methods and identify their limitations in modeling confined fluid compressibility in large pores. The isothermal compressibility $\beta_T$, and its reciprocal, isothermal bulk modulus $K_T$, are defined as
\begin{equation}
    K_{T} = \beta_{T}^{-1} = -V \left( \frac{\partial P}{\partial V} \right)_{T},
    \label{eq:modulus_thermo}
\end{equation}
where $V$ is the volume of the system, $P$ is the fluid pressure, and the temperature $T$ is held constant. This macroscopic definition of bulk modulus assumes that the system pressure is isotropic. 

Through statistical mechanics, derivative thermodynamic properties of a system can be calculated from the fluctuation of extensive variables in a given ensemble. In the grand canonical ensemble, $V$, $T$, and chemical potential $\mu$ are prescribed and held constant, while the number of particles $N$ in the system is allowed to vary, most commonly by Monte Carlo moves. The bulk modulus of such a system can be calculated through the expression
\begin{equation}
    K_{T} = \frac{k_{\rm B} T \langle N\rangle^2}{V \langle \delta N^2 \rangle},
    \label{eq:N_fluct}
\end{equation}
where $k_{\rm B}$ is the Boltzmann constant, angled brackets represent an ensemble average, and $\langle \delta N^2 \rangle = \langle N^2 \rangle - \langle N \rangle^2$ is the variance of $N$. In the canonical (NVT) ensemble, $N$, $V$, and $T$ are held constant. This leaves fluctuations in the internal virial $\mathcal{W}$ and hypervirial function $\mathcal{X}$, from which the bulk modulus can also be calculated as
\begin{equation}
    K_{T} = \frac{1}{V} \left( N k_{\rm B} T + \langle \mathcal{W} \rangle + \langle \mathcal{X} \rangle - \frac{\langle \delta \mathcal{W}^2 \rangle}{k_{\rm B}T} \right)
    \label{eq:virial_fluct}
\end{equation}
in which $\langle \delta \mathcal{W}^2 \rangle$ is the variance of the virial, defined similarly to the variance of $N$ in Eq.~\ref{eq:N_fluct}~\cite{Allen2017}. 

These methods of modeling bulk modulus, while effective for simulations of simple fluids confined in a relatively small pore, are not easily applied to all fluids or to a large pore. In GCMC simulations of dense fluids, it is difficult to reach and sample an equilibrated system in a reasonable amount of time because particle insertion becomes progressively inefficient as the size of the system increases. This approach is further hampered by the introduction of Coulombic and many-body interactions. In contrast, molecular dynamics (MD) simulations of simple fluids in the NVT ensemble are relatively fast, but the calculations required to apply the virial fluctuation method add further computational cost. Additionally, it is not applicable for molecules with flexible angles and dihedrals.

A method which has neither of the aforementioned limitations can be applied in the isothermal-isobaric (NPT) ensemble, in which $N$, $P$, and $T$ are held constant, while the volume of the system fluctuates. In such a system, the bulk modulus can be calculated as
\begin{equation}
    K_{T} = \frac{k_{\rm B} T \langle V\rangle}{\langle \delta V^2 \rangle},
    \label{eq:V_fluct}
\end{equation}
where $\langle \delta V^2 \rangle$ is the variance of $V$, defined similarly to the variance of $N$ given above. This method is routinely applied to bulk fluids and solids. However, the application of Eq.~\ref{eq:V_fluct} to fluids confined by solid pore walls is complicated by the following challenges. (1) Isotropic volume fluctuations will change the pore size, and thus the bulk modulus of the fluid during the course of the simulation. (2) Even if volume fluctuations are constrained so that the pore size is constant, it would require the real-time addition and removal of atoms making up the solid walls to keep the fluid confined along the now fluctuating edges of the pore volume. (3) Although the atoms comprising the solid walls can be ``frozen'' and do not contribute to barostat calculations, the total volume of the simulation box (and thus the variance of that quantity) will include the contribution of the pore walls, which must be removed. To access the performance benefits of using Eq.~\ref{eq:V_fluct}, the method proposed herein avoids the aforementioned issues by using infinite integrated potentials to represent the confining walls and allowing the volume to fluctuate only in the remaining periodic dimensions.

\section{Integrated Potentials}

Integrated potentials are commonly used in Monte Carlo and MD simulations to modify the potential landscape experienced by atoms in the system, primarily to save computational time~\cite{Siderius2011, Maximov2021}. For example, a flat wall of frozen atoms interacting with a fluid can be replaced by a single immaterial plane which imposes a force on a fluid particle as a function of that particle's distance from the plane. This potential has approximately the same effect on the fluid particles as would a physical wall without calculating the interaction potential contributed by every wall atom. They are especially useful in classical density functional theory (DFT) calculations~\cite{Landers2013}.

\begin{figure}[t]
    \centering
    \subfloat[]{\includegraphics[width=0.9\textwidth]{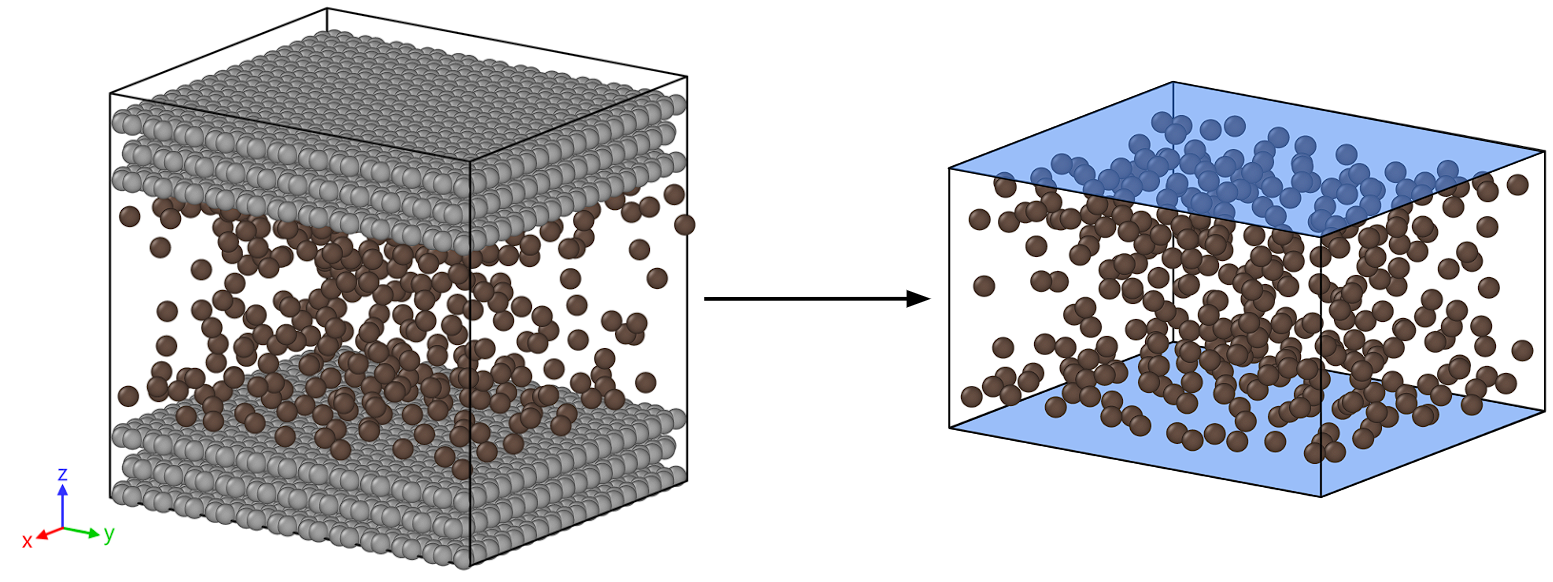}}\\
    \subfloat[]{\includegraphics[width=0.85\textwidth]{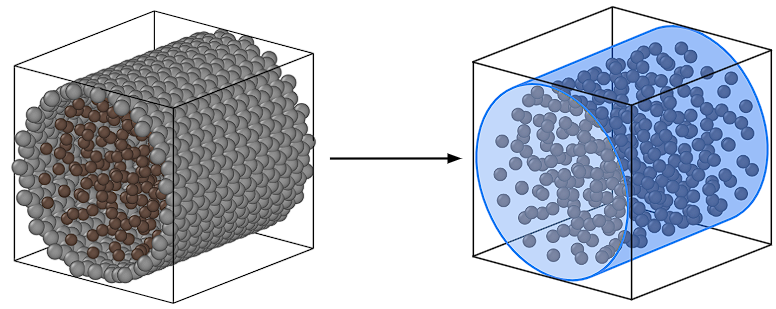}}\\
    \caption{\footnotesize{Schematic simulation snapshots of methane confined in \SI{3}{nm} (a) slit and (b) cylindrical pores with atomistic walls (left) and integrated potential walls (right)}}
    \label{fig:Snapshots}
\end{figure}

To determine the most accurate form of the potential, we first account for the contribution of each wall atom. We use as a starting point the pair-wise 12-6 Lennard-Jones (LJ) interaction potential commonly used in molecular simulations of non-polar fluids, which quantitatively reproduces the bulk modulus of methane~\cite{Corrente2020}:
\begin{equation}
    U = 4\epsilon \left[ \left( \frac{\sigma}{r}\right)^{12} -\left( \frac{\sigma}{r}\right)^6 \right].
    \label{eq:Lennard-Jones_12-6}
\end{equation}
Here $U$ is the potential energy applied to the two interacting particles, $\epsilon$ and $\sigma$ are respectively the energy and distance parameters characterizing the interaction, and $r$ is the distance between the two particles. Interactions are normally truncated at some distance $r_{\rm cut}$ so that the potential between particles further than this distance is set to zero. Typically, pore walls are represented by a number of stationary atoms, each of which interacts with fluid particles according to Eq.~\ref{eq:Lennard-Jones_12-6}, and are arranged in the desired pore geometry. In this work, slit pores are represented by two infinite planes between which the fluid is confined, and cylindrical pores are represented by an infinite cylindrical shell (see Fig.~\ref{fig:Snapshots}).

The potential energy landscape experienced by a fluid atom due to flat walls is obtained by twice-integrating Eq.~\ref{eq:Lennard-Jones_12-6} along the two dimensions with which the wall is parallel, yielding:
\begin{equation}
    U_{\rm sf} = \sum_{i=1}^{n}{ \left[ 4 \pi \rho_{\rm s} \epsilon_{\rm sf} \sigma_{\rm sf}^2 \left( \frac{1}{5} \left[ \frac{\sigma_{\rm sf}}{r_{i}} \right]^{10} - \frac{1}{2} \left[ \frac{\sigma_{\rm sf}}{r_{i}} \right]^4 \right) \right]},
    \label{eq:Lennard-Jones_10-4}
\end{equation}
where $\rho_{\rm s}$ is the number density of LJ sites (solid atoms) in each wall layer, the subscript ``sf" denotes interaction parameters as in Eq.~\ref{eq:Lennard-Jones_12-6} between the solid and fluid atoms, $n$ is the total number of layers in both walls (such as graphene layers in graphitic walls), and $r_{i}$ is the distance between a fluid atom and the $i^{\rm th}$ wall layer.

The potential energy landscape experienced by a fluid atom confined in a cylindrical pore is obtained by integrating Eq.~\ref{eq:Lennard-Jones_12-6} in cylindrical coordinates, yielding the potential developed by Tjatjopoulos et al.~\cite{Tjatjopoulos1988}:
\begin{equation}
    U_{\rm sf}(r,R) = 2 \pi \rho_{\rm s} \sigma_{\rm sf}^2 \epsilon_{\rm sf} [\psi_6(r,R,\sigma_{\rm sf}) - \psi_3(r,R,\sigma_{\rm sf})],
    \label{eq:Tjatjopoulos}
\end{equation}
where $r$ is the distance between the fluid atom and the central axis parallel to the cylinder, $R$ is the cylinder radius, and $\psi_{n}$ is a function defined in Eq. \ref{eq:SI_psi_n} of Supplemental Material. For more comprehensive details the reader is directed to Refs.~\cite{Tjatjopoulos1988, Siderius2011}.

\section{LAMMPS Implementation}

We implemented this method for both slit and cylindrical pores in LAMMPS (Large Atomic and Molecular Massively Parallel Simulator)~\cite{LAMMPS}, which has built-in support for integrated potentials. Slit pore walls were created using a custom command similar to \texttt{fix wall/lj1043}, but modified to impose the potential defined in Eq.~\ref{eq:Lennard-Jones_10-4}. Two planes were defined to create the slit pore: one at both ends of the simulation box in the z-direction. Different pore sizes were modeled by modifying the height of the simulation box. To test that Eq.~\ref{eq:Lennard-Jones_10-4} is an accurate representation of slit pore confinement, we compared the fluid density profiles in a \SI{3}{nm} slit pore using both atomistic walls and the integrated potential, included in Supplemental Material (Figure~\ref{fig:SI_Potentials_and_density}).

Modeling cylindrical pores in LAMMPS first required the creation of the cylindrical region onto which the wall potential would be mapped. None of the wall potentials currently available in LAMMPS adequately model the landscape described by Eq.~\ref{eq:Tjatjopoulos}. To remedy this, it was necessary to write the potential explicitly as an additional option. The Tjatjopoulos potential was then applied to this region using the \texttt{fix wall/region} command. For the full syntax used to create slit and cylindrical pore walls, see Supplemental Material.

\section{Results}

\subsection{Methane in Carbon Slit and Cylindrical Pores}

To demonstrate the proposed method, we chose to simulate the system of supercritical methane (represented by a united-atom model~\cite{Martin1998}) confined in graphitic slit pores, which has practical importance as a proxy for modeling shale gas or coal bed methane~\cite{Mosher2013}. This same system was studied by Corrente et al.~\cite{Corrente2020}, in which the authors used both Eq.~\ref{eq:N_fluct} and Eq.~\ref{eq:virial_fluct} to calculate the bulk modulus of the confined fluid. Calculations were carried out for pore sizes ranging from \SI{3}{nm} to \SI{9}{nm} and reservoir pressures from \SI{2}{MPa} to \SI{10}{MPa}. It is worth noting that the bulk modulus as calculated by Eq.~\ref{eq:V_fluct} is sensitive to the exact number of fluid atoms in the pore, which was determined by GCMC simulations. To extend this method to larger pore sizes without running lengthy GCMC simulations, it was necessary to extrapolate the number of atoms as described later in this letter. Furthermore, calculating the bulk modulus from Eq.~\ref{eq:V_fluct} for larger pores may introduce some finite-size effects. An analysis of these effects on the modulus is given in Supplemental Material.

\begin{figure}[t]
     \centering
     \subfloat[\chem{CH_4} in Slit Pores]{\includegraphics[width=0.5\textwidth]{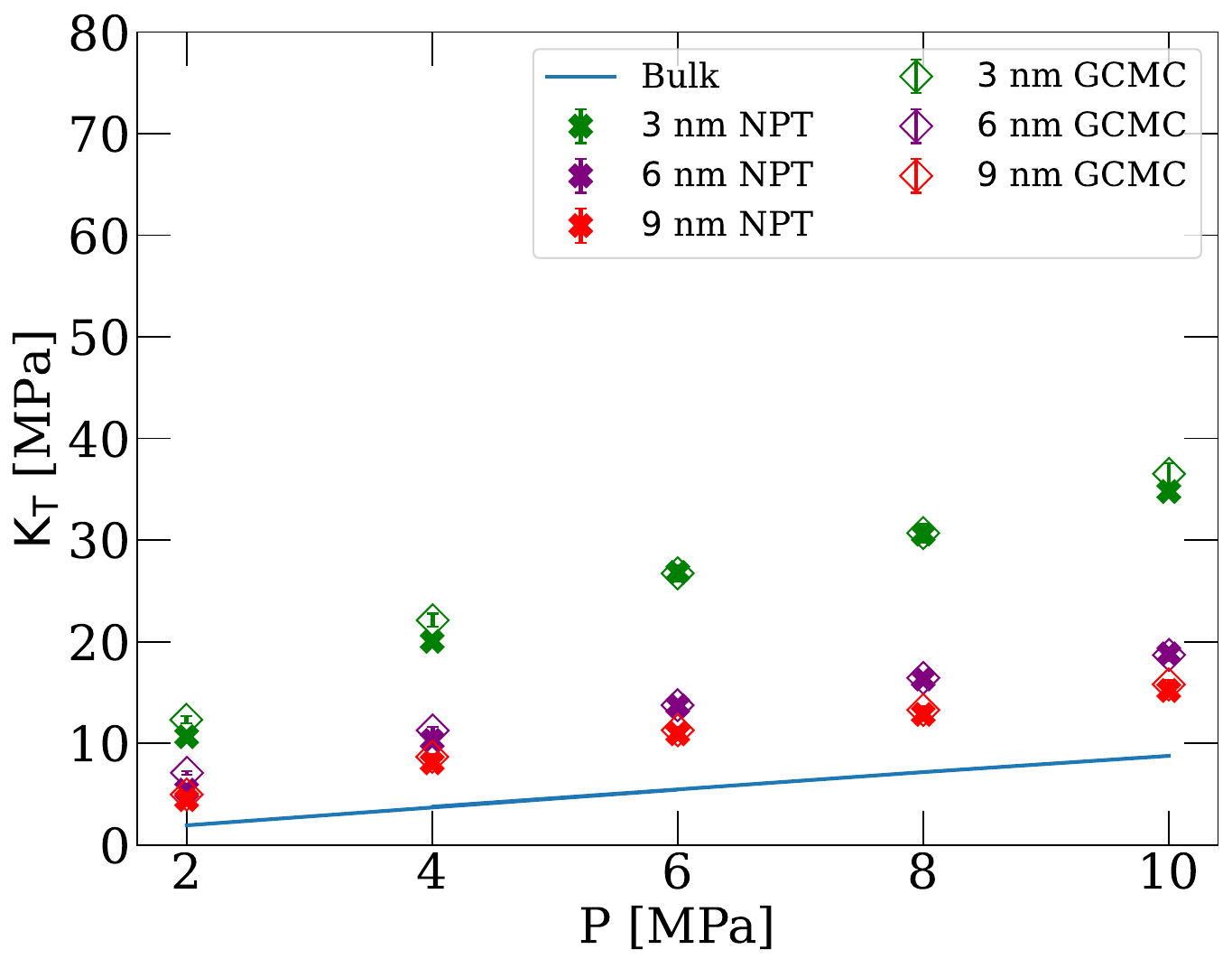}}
     \subfloat[\chem{CH_4} in Cylindrical Pores]{\includegraphics[width=0.5\textwidth]{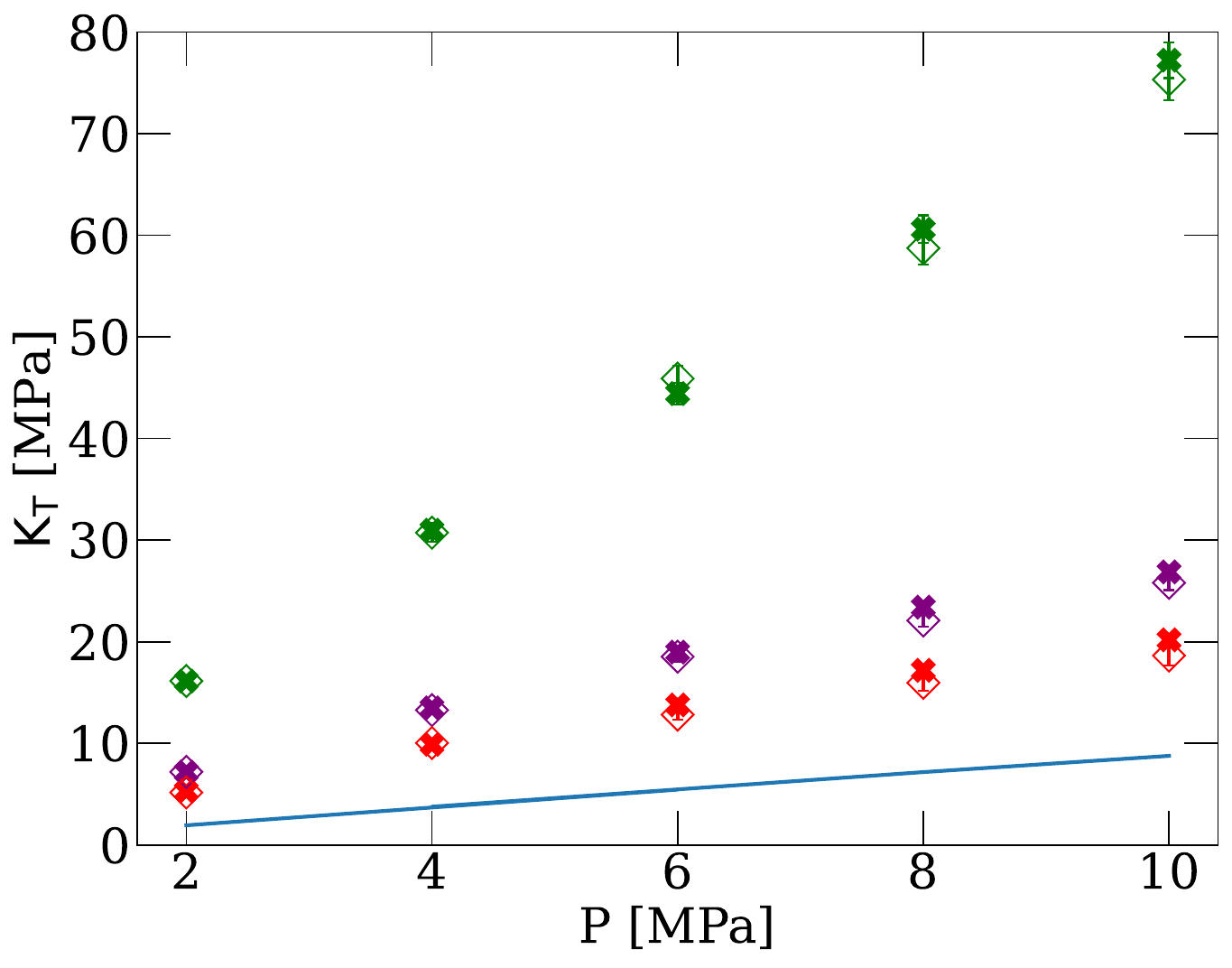}}
     \caption{\footnotesize{Bulk modulus of supercritical methane as a function of reservoir pressure in (a) carbon slit and (b) cylindrical pores. The filled ``$\times$" markers are the results calculated using the new method, from Eq.~\ref{eq:V_fluct}. The empty markers represent results calculated from Eq.~\ref{eq:N_fluct} (diamonds). The solid line is the modulus of bulk methane. In (a), the data are taken from Ref.~\cite{Corrente2020}.}}
     % \caption{\footnotesize{Bulk modulus of supercritical methane as a function of reservoir pressure in (a) carbon slit and (b) cylindrical pores, respectively. The filled ``X" markers are the results calculated using the new method, from Eq.~\ref{eq:V_fluct}. The empty markers represent results calculated from Eq.~\ref{eq:N_fluct} (diamonds) or Eq.~\ref{eq:virial_fluct} (squares). In (a), those data are taken from Ref.~\cite{Corrente2020}.}}
     \label{fig:K_vs_P}
\end{figure}

The calculated bulk moduli are shown in Fig.~\ref{fig:K_vs_P}a, compared with GCMC results. Across all pore sizes and reservoir pressures, we observed quantitative agreement between Eq.~\ref{eq:N_fluct} and the method introduced in this letter. Details regarding the calculation of statistical errors can be found in Supplemental Material.

Having demonstrated that our method is effective for slit pores, we extend it to another geometry by testing the same solid-fluid pair but in cylindrical pores. As with slit pores, GCMC simulations were employed to calculate the mean density of adsorbed fluid at equilibrium, which was then used to generate the initial configurations for MD simulations in LAMMPS. Eqs.~\ref{eq:Lennard-Jones_10-4} and \ref{eq:Tjatjopoulos} both used solid-fluid interaction parameters calculated via conventional Lorentz-Berthelot mixing rules~\cite{Lorentz1881}. All interaction parameters can be found in Table \ref{tab:SI_LJ_Parameters} of Supplemental Material.

The bulk moduli as computed through Eqs.~\ref{eq:N_fluct} and ~\ref{eq:V_fluct} are shown in Fig.~\ref{fig:K_vs_P}b in pore sizes mirroring the results for slits. As with slit pores, agreement is observed between the bulk moduli calculated from both methods. For both pore shapes, the modulus shows a monotonic increase with the pressure, similarly to bulk fluids. However, the modulus exceeds that of the bulk -- the fluid in the pores appears much stiffer.

\subsection{Bulk Modulus vs Inverse Pore Size}

\begin{figure}[t]
    \centering
    \subfloat[\chem{CH_4} in Slit Pores]{\includegraphics[width=0.45\textwidth]{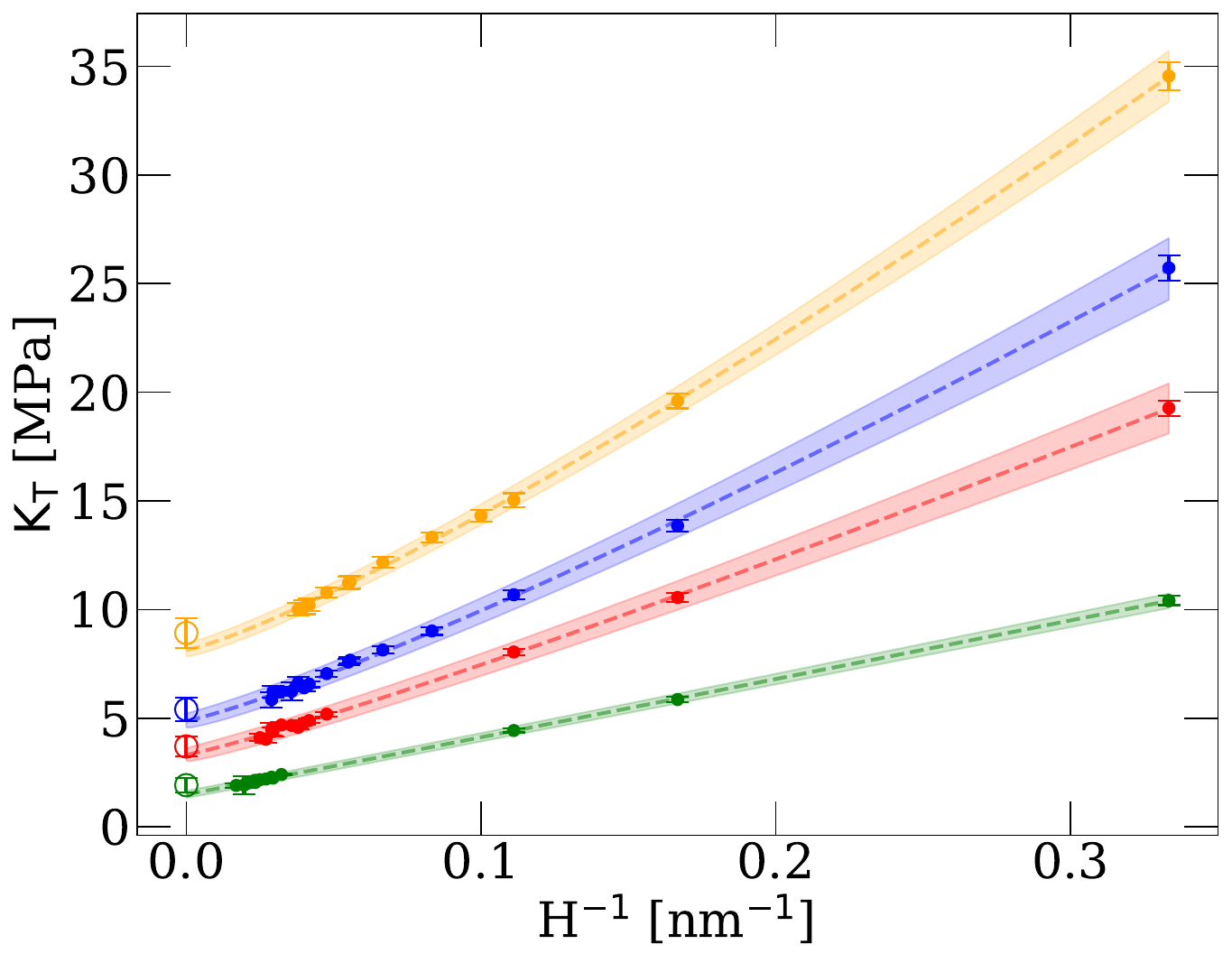}}
    \subfloat[\chem{CH_4} in Cylindrical Pores]{\includegraphics[width=0.45\textwidth]{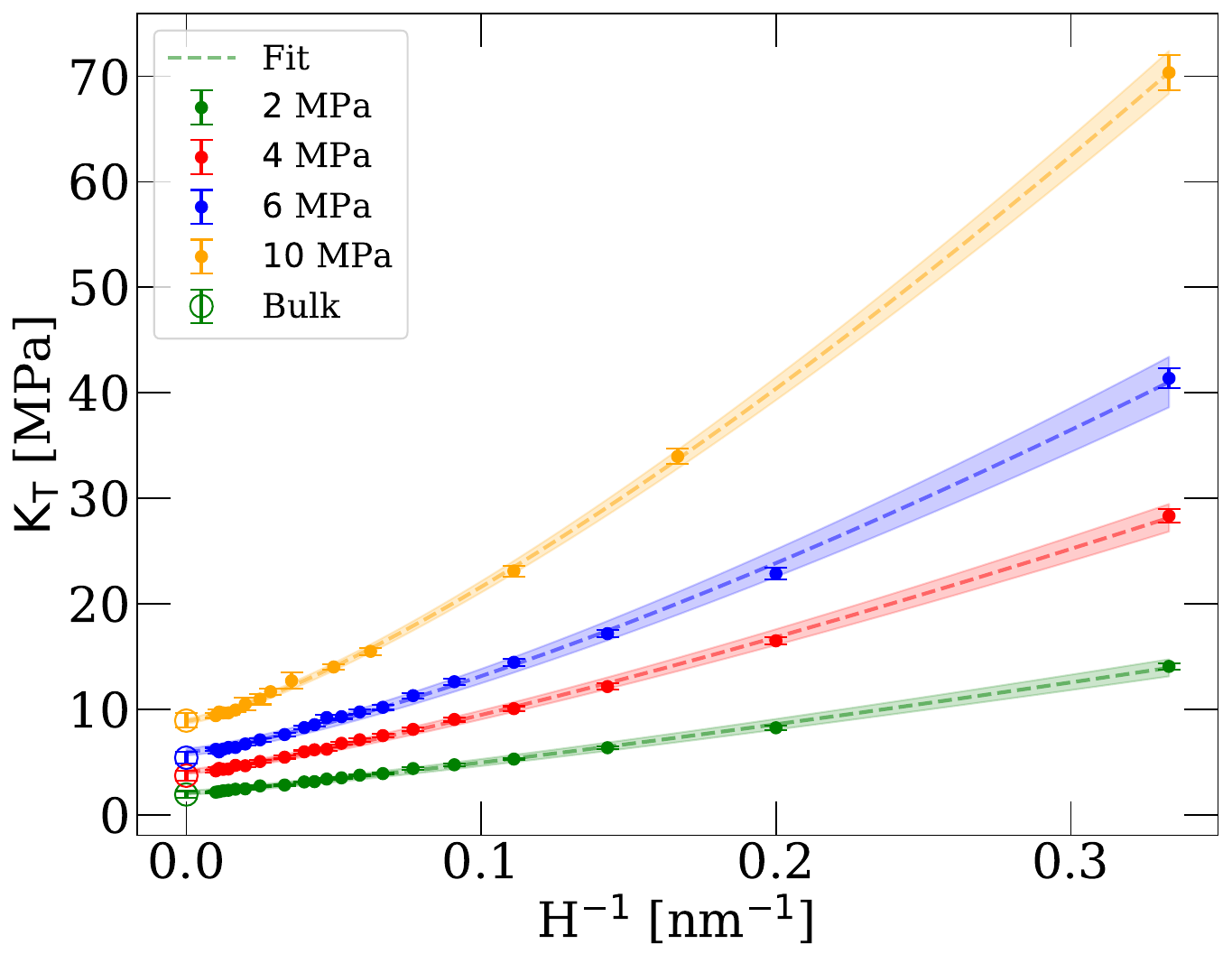}} \\
    \subfloat[Close-up of (a)]{\includegraphics[width=0.45\textwidth]{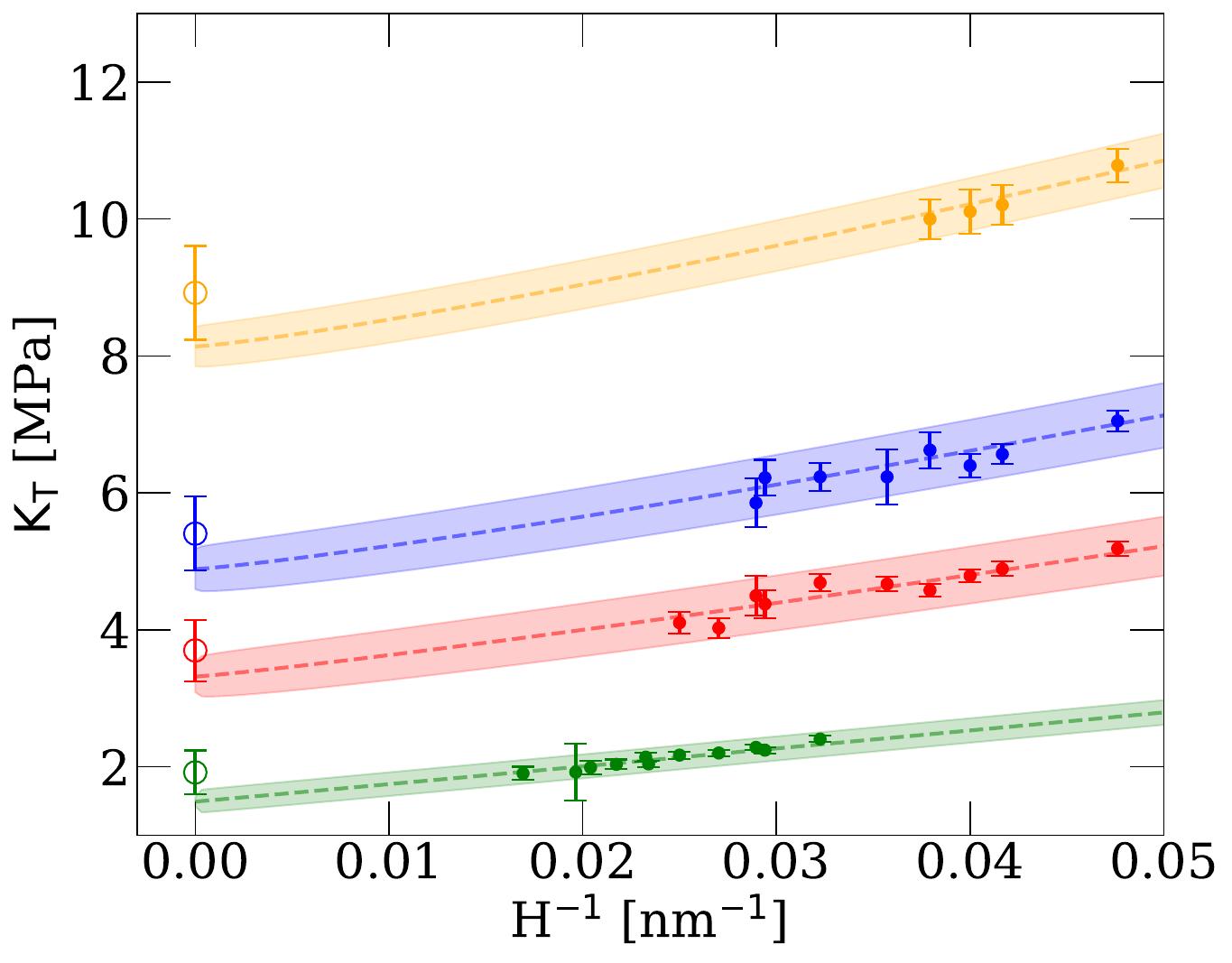}}
    \subfloat[Close-up of (b)]{\includegraphics[width=0.45\textwidth]{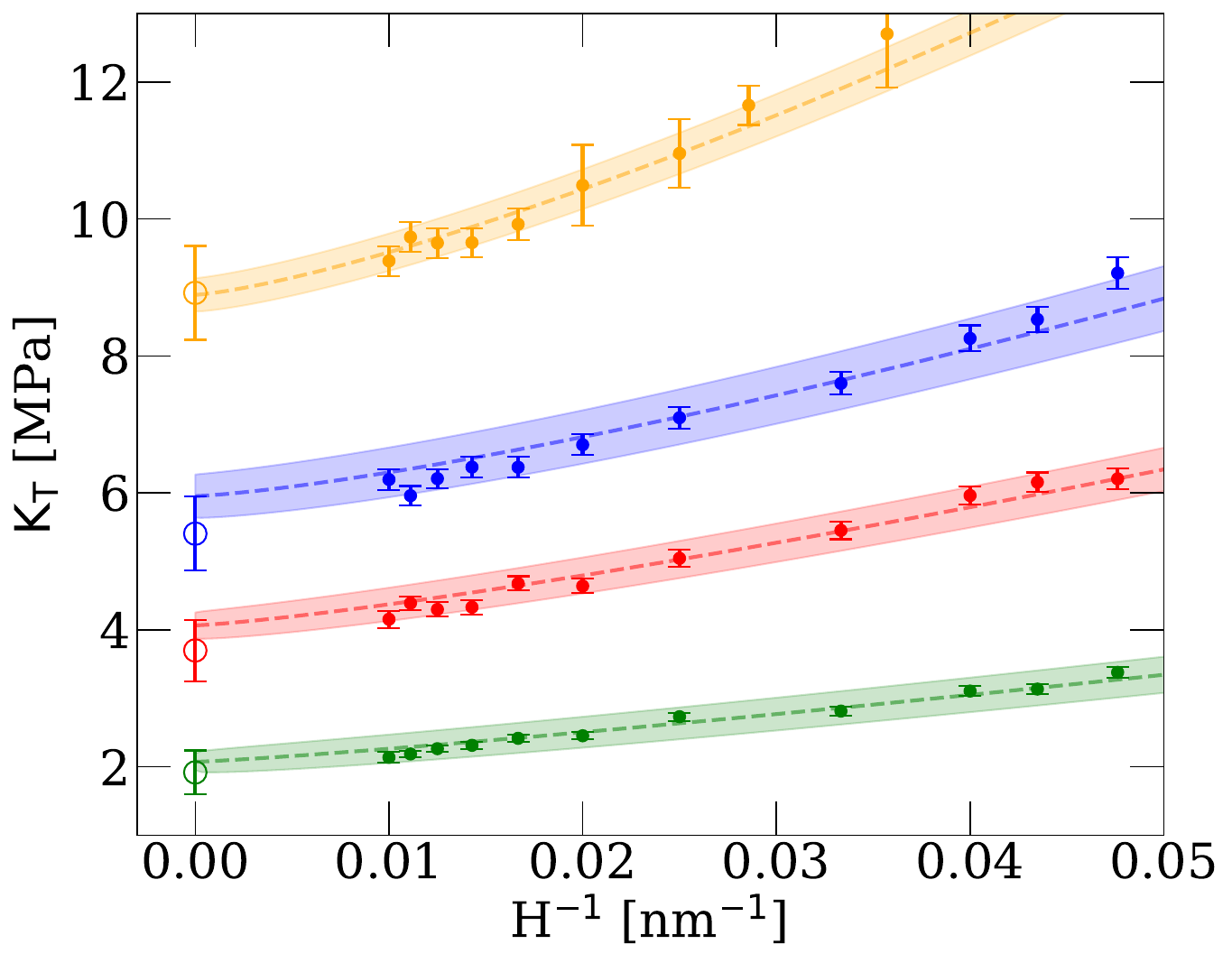}}
    \caption{\footnotesize{Bulk modulus of supercritical methane vs inverse pore size at reservoir pressures of $\SI{2}{MPa}-\SI{10}{MPa}$ in slit pores (a) and (c), and cylindrical pores (b) and (d). The empty circles represent the bulk moduli of non-confined methane at the corresponding reservoir pressure. The dashed lines are a curve fit to $K_T({H}^{-1}; a,b,c) = a ({H}^{-1})^b + c$, and the shaded areas are the error of the fit.}}
    \label{fig:K_vs_invH}
\end{figure}

Finally, we apply the new method to investigate the fluid bulk modulus in pore sizes up to \SI{100}{nm} in slit and cylindrical pores to demonstrate the gradual transition from confined to bulk-like behavior, illustrated in Fig.~\ref{fig:K_vs_invH}. Corrente et al.~\cite{Corrente2020} showed the bulk modulus of methane in carbon slit pores of 2 to \SI{9}{nm} is linear with respect to the inverse pore size below the critical pressure (\SI{4.6}{MPa} at \SI{298}{K}). Alongside the data in Fig.~\ref{fig:K_vs_invH}, we plot corresponding best-fits for the function $K = a H^{-b} + c$ as dashed lines. This demonstrates the gradual approach towards the bulk limit, with higher pressures exhibiting stronger non-linear behavior than lower pressures. The same behavior is observed in cylindrical pores. Even above the critical point, we observe largely linear behavior in smaller pores (less than \SI{10}{nm}), with notable deviations from linearity only appearing towards the limit of a large pore.

When a fluid is confined in nanopores, its pressure varies in space according to the confining geometry and the distance from the pore walls. This means that the bulk modulus of a confined fluid is also spatially dependent~\cite{Sun2019density, Dobrzanski2021}. However, when considering the mechanical response of a nanoporous solid saturated with fluid, probed via ultrasound, one generally observes the averaged properties of the fluid, and does not observe spatial variations on the scale of a single pore~\cite{Dobrzanski2021}. We focus accordingly on the average bulk modulus of the fluid in a pore. It is important to note that, although Fig.~\ref{fig:K_vs_invH} shows a slight underestimation of the fitted modulus for large slit pores, the predicted moduli from our new method agree with the moduli of bulk methane, as shown in Fig.~\ref{fig:SI_KT_vs_Lx} of Supplemental Material.

We have demonstrated a simulation technique that applies fluctuation theory to accurately model the bulk modulus of a confined Lennard-Jones fluid in a wide range of pore sizes. Further, the parameters that define the interaction between the wall and fluid can easily be modified to represent other solid-fluid combinations. Integrated potentials can conceivably be applied for pore geometries other than idealized slits and cylinders, such as wedge pores and prisms with polygonal cross-sections. Conceptually, the only requirement is the presence of at least one periodic dimension along which the fluid volume is allowed to fluctuate in a constant-stress ensemble. While this limitation prevents direct application to spherical pores, it may be possible to model them empirically by simulating slits and cylinders of various sizes and seeking to describe the combined effect of curvature and solid-fluid interactions on the fluid bulk modulus using the corresponding states principle for confined fluids~\cite{An2023}. This method enables such a systematic approach, and is especially useful for complex fluids for which Monte Carlo poses a hard limitation, such as longer alkanes. 

\section{Computational Methods} \label{Methods}

\subsection{Initial Configurations for Large Pores} \label{Extrapolation}

Initial configurations were generated by assigning random positions of methane molecules in the pore. Then, the LAMMPS default energy minimization technique was applied to avoid any molecule-molecule or molecule-wall overlap. To determine the number of molecules to be inserted, we reproduced adsorption of methane in cylindrical and slit pores using GCMC simulations (from \SI{3}{nm} to \SI{10}{nm} in size) and fitted the resulting excess adsorbed amount, $N_{\rm ex}=(\rho_{\rm f}-\rho_{\rm b})V_{\rm f}$, according to the Gamma distribution:
\begin{equation}
\label{eq:Nex}
    N_{\rm ex}(V_{\rm f}; B, \alpha, \theta) = B\frac{V_{\rm f}^{\alpha-1}e^{-V_{\rm f}/\theta}}{\theta^{\alpha}\Gamma{(\alpha)}},
\end{equation}
where $B$, $\alpha$, and $\theta$ are the fitting parameters, and $V_{\rm f}$ is the volume of the fluid in the pore ($V_{\rm f}=H_{\rm int}\times A$ for slits and $V_{\rm f}=\pi[H_{\rm int}^2/4]\times 3r_{\rm cut}$ for cylinders). $H_{\rm int}=H-1.7168\sigma_{\rm sf}+\sigma_{\rm ff}$ is the internal pore size and $A$ is the area of a pore wall~\cite{Dobrzanski2018}. $r_{\rm cut}$ is the cutoff radius of solid-fluid interactions. The shape parameter, $\alpha$, is adjusted according to the pore-shape ($\alpha\approx1.5$ for cylindrical pores and $\alpha\approx1$ for slit pores), while the scale and magnitude parameters, $\theta$ and $B$, are dependent on the thermodynamic conditions. Extrapolating the adsorbed amount of methane allows us to avoid running lengthy GCMC simulations of large systems (pores larger than \SI{10}{nm} in this work) and save significant computational time. Validation of Eq. \ref{eq:Nex} can be found in Supplemental Material.

The resulting adsorbed amount is:
\begin{equation}
    N = \rho_{\rm b}V_{\rm f} + N_{\rm ex},
\end{equation}
where $\rho_{\rm b}$ is the density of bulk methane, and $N_{\rm ex}$ comes from Eq. \ref{eq:Nex}. $N_{\rm ex}$ obtained from the Gamma distribution is strictly positive, and decreases with increasing pore size. For sufficiently large pores, the dominant term will be $\rho_{\rm b}V_{\rm f}$. Therefore, the density of confined methane will approach the density of bulk methane.

To reduce any possible finite-size effects in volume oscillations, we set initial wall areas of $A=H\times H$ for slit pores and $A=\pi H\times 3r_{\rm cut}$ for cylindrical pores. Figure \ref{fig:K_vs_P} shows that increasing $A$ even further would show a negligible improvement of $K_T$ predictions from MD simulations.

\subsection{Simulation Procedure} \label{Simulation_Procedure}

MD simulations in slit pores began with equilibration in the canonical ensemble for \num{1e6} timesteps (\num{2e6} in cylinders) with one timestep set to \SI{0.1}{fs}. During this phase, the normal components of the system's pressure tensor ($P_{\rm xx}$, $P_{\rm yy}$, $P_{\rm zz}$) were averaged using the \texttt{compute pressure} and \texttt{fix ave/time} commands. The ensemble was then switched from NVT to NPT, in which the calculated pressure values are passed to the Nose-Hoover barostat and volume fluctuations are measured for \num{2e7} timesteps.

For slit pores, only the pressure in the x- and y-directions (those directions which are parallel to the pore walls) are specified in the \texttt{fix npt} command with the \texttt{aniso} option for the barostat, thus allowing the volume of the fluid to fluctuate without changing the pore size. Both the x- and y- components specified in the \texttt{fix npt} command use the same value: the arithmetic mean between $\langle P_{\rm xx} \rangle$ and $\langle P_{\rm yy} \rangle$. In cylindrical pores, only $\langle P_{\rm xx} \rangle$ is given to the barostat, thus allowing fluctuation only in the x-direction. The normal components of the pressure tensor had to be measured over the course of the NVT simulation because the pressure in the pore is known to differ from that of the bulk reservoir with which the pore fluid is in equilibrium (i.e., it would be physically inaccurate to apply a pressure of \SI{10}{MPa} to methane confined in a \SI{3}{nm} pore based on the assumption that the fluid pressure is equal to the reservoir pressure). We also observed that the pressures in the periodic dimensions gradually approached the pressure of the reservoir with increasing pore size.

Over the course of the NPT simulation, volume was collected once every $10^3$ timesteps, and the first \num{5e6} steps were skipped to allow volume to equilibrate before calculating the bulk modulus through Eq.~\ref{eq:V_fluct}. In both slit and cylindrical pores, the volume used was the fluid-accessible pore volume ($V_{\rm f}$ in Eq.~\ref{eq:Nex}). Slit pores used 3 graphene layers in each wall ($n = 3$ in Eq.~\ref{eq:Lennard-Jones_10-4}) and cylindrical pores were single-layered.

\begin{acknowledgments}

This work used the supercomputers Engaging at the Massachusetts Green High Performance Computing Center, Wulver at the New Jersey Institute of Technology, and DARWIN at the University of Delaware. The latter through the allocation CHM240028 from the Advanced Cyberinfrastructure Coordination Ecosystem: Services \& Support (ACCESS) program, which is supported by National Science Foundation grants \#2138259, \#2138286, \#2138307, \#2137603, and \#2138296~\cite{Boerner2023}. This work was supported by the National Science Foundation (grant CBET-2344923).

\end{acknowledgments}

\section*{Supplemental Material}

See Supplemental Material for information about the integrated potentials for slit and cylindrical pores, LAMMPS details about the implementation of the integrated potentials and interaction parameters, validation of the particle count extrapolation to large pores, estimation of statistical errors, fitting parameters for the bulk modulus as a function of the inverse pore size, and an analysis of the finite-size effects involved in calculating $K_T$ from Eq.~\ref{eq:V_fluct}. ~\href{run:./Supplementary.tex}{PDF}.

\newpage

\begin{center}
\section*{Supplemental Material}
\end{center}

\appendix

\renewcommand{\theequation}{S\arabic{equation}}
\renewcommand{\thetable}{S\arabic{table}}
\renewcommand{\thefigure}{S\arabic{figure}}
\setcounter{figure}{0}

\subsection{LAMMPS Details} \label{SI_LAMMPS}

The exact syntax used to initialize the LJ walls representing slit pores was: \texttt{fix f\_ID g\_ID wall/lj104 zlo EDGE \$\{eps\_sf\} \$\{sig\_sf\} \$\{wall\_cut\} \$\{rho\_A\} \$\{nLayers\} \$\{deltaLayers\} zhi EDGE \$\{eps\_sf\} \$\{sig\_sf\} \$\{wall\_cut\} \$\{rho\_A\} \$\{nLayers\} \\ \$\{deltaLayers\}}, where \texttt{f\_ID} is the fix ID, \texttt{g\_ID} is the ID of the group of atoms interacting with the wall (fluid atoms), \texttt{\$\{eps\_sf\}} and \texttt{\$\{sig\_sf\}} are the fitted interaction parameters, \texttt{\$\{wall\_cut\}} is the cutoff distance beyond which atoms do not interact with the wall, \texttt{\$\{rho\_A\}} is the surface density of atoms in the pore wall, and \texttt{\$\{nLayers\}} and \texttt{\$\{deltaLayers\}} are the number of layers in the wall and the distance between those layers, respectively. The specification of a group ID other than \texttt{all} allows for mixtures to be simulated by overlaying walls on top of each other, each wall interacting with a different atom type through unique parameters. The walls take up the entire edge of the simulation box, even as the area of the edge changes during the NPT section of the procedure.

The exact syntax used to initialize the integrated potential representing cylindrical pores was: \texttt{fix f\_ID g\_ID r\_ID wall/region/tjatjopoulos \$\{eps\_sf\} \$\{sig\_sf\} \$\{rho\_A\} \$\{size\}}, where \texttt{wall/region/tjatjopoulos} is the custom fix style and \texttt{\$\{size\}} is the diameter of the pore. \texttt{r\_ID} is the region ID, which sets the cylindrical region of the pore. The region was defined as: \texttt{region r\_ID cylinder x 0 0 \$(size/2) INF INF side in}. This places the cylinder in the center of the box, with its central axis parallel to the x-axis. The cylinder is infinite in the axial direction so that the potential surface persists even as the length of the box in that direction increases.

\begin{table*}[b]
\caption{\footnotesize{Lennard-Jones (LJ) interaction parameters for the simulations performed in this work. The ``\chem{C} - \chem{CH_4}" parameters were calculated using Lorentz-Berthelot mixing rules~\cite{Lorentz1881} and are the parameters used in Eq.~\ref{eq:SI_Lennard-Jones_10-4} and in the potential by Tjatjopoulos. $r_{\rm cut}$ is the distance beyond which LJ interactions are truncated and set to zero.}}
\label{tab:SI_LJ_Parameters}
\begin{ruledtabular}
\begin{tabular}{lccc}
Interaction & $\epsilon$ (kcal/mol) & $\sigma$ (\AA) & $r_{\rm cut}$ (\AA) \\
\colrule
\chem{C} - \chem{C} & 0.0556 & 3.40 & - \\
\chem{CH_4} - \chem{CH_4} & 0.2941 & 3.73 & 12.0 \\
\chem{C} - \chem{CH_4} & 0.128 & 3.57 & 12.0 \\
\end{tabular}
\end{ruledtabular}
\end{table*}

\subsection{Potential by Tjatjopoulos et al.} \label{sec:SI_Tjatjopoulos}

The potential used to represent cylindrical pores was originally defined in Ref.~\cite{Tjatjopoulos1988}. Its implementation in LAMMPS is based on the equations derived in Ref.~\cite{Siderius2011}. The $\psi_{\rm n}$ functions in the main text are defined as follows:
\begin{equation} \label{eq:SI_psi_n}
    \psi_{\rm n}(r,R,\sigma) = 4 \sqrt{\pi} \frac{\Gamma(n - \frac{1}{2})}{\Gamma(n)}\left(\frac{\sigma}{R}\right)^{2n-2} \left[ 1 - \left(\frac{r}{R}\right)^2 \right]^{2-2n} F\left[\frac{3-2n}{2},\frac{3-2n}{2};1;\left(\frac{r}{R}\right)^2\right]
\end{equation}
where $\Gamma$ is the Gamma function, $F$ is the hypergeometric function, and $n$ is an integer or half-integer greater than one half.

\subsection{Validation of Integrated Potential} \label{sec:SI_LJ10-4}

\begin{figure}[t]
    \centering
    \subfloat[]{\includegraphics[width=0.49\textwidth]{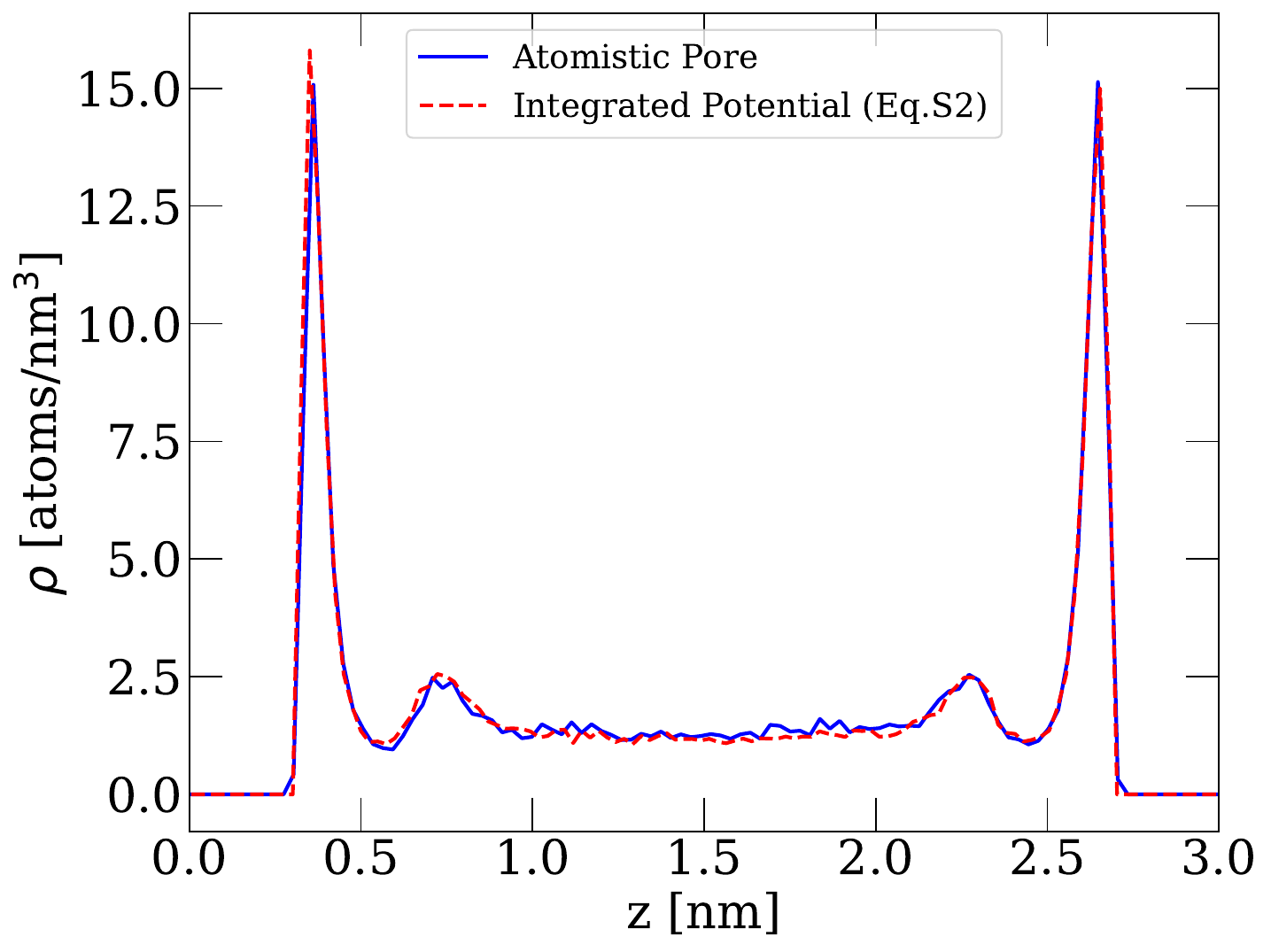}}
    \caption{\footnotesize{Density profiles of methane in a \SI{3}{nm} carbon slit pore. The blue line represents the fluid density in a physical, atomistic pore. The red dashed line represents the fluid confined between integrated potential walls defined by Eq.~\ref{eq:SI_Lennard-Jones_10-4}}}
    \label{fig:SI_Potentials_and_density}
\end{figure}

To ensure that the integrated potential walls are accurate representations of a pore, the fluid density profiles across the height of the pore were compared. Two simulations of confined united-atom methane~\cite{Martin1998} at $T$ = \SI{298}{K} were performed, once with all-atom graphitic walls and once with integrated potentials defined by:
\begin{equation}
    U_{\rm sf} = \sum_{i=1}^{n}{ \left[ 4 \pi \rho_{\rm s} \epsilon_{\rm sf} \sigma_{\rm sf}^2 \left( \frac{1}{5} \left[ \frac{\sigma_{\rm sf}}{r_{i}} \right]^{10} - \frac{1}{2} \left[ \frac{\sigma_{\rm sf}}{r_{i}} \right]^4 \right) \right]},
    \label{eq:SI_Lennard-Jones_10-4}
\end{equation}
as described in the main text. See Table~\ref{tab:SI_LJ_Parameters} for the LJ parameters governing methane-methane and methane-wall interactions. Both simulations used the same pore area (\SI{4.272}{nm} $\times$ \SI{4.932}{nm} as in Ref.~\cite{Corrente2020}) and pore size (\SI{3}{nm}). Figure~\ref{fig:SI_Potentials_and_density} compares the resultant density profiles of the confined fluid. The agreement between density profiles gives some assurance that the fluid should exhibit some of the same physical behavior in this simplified pore as it would in an all-atom pore.

\subsection{Extrapolation of Particle Count in Large Pores} \label{sec:SI_extrap}

The method proposed in the main text requires an initial density to populate the simulation box. One approach to obtain the initial density is from GCMC simulations. As long as we know $\mu=\mu(P)$ (for example, from an equation of state), we can predict the density of the fluid in a pore using GCMC simulations and then compute the bulk modulus from MD simulations. However, simulating large systems using GCMC can be lengthy. To avoid running additional GCMC simulations, we propose confined fluid density in small pores using GCMC simulations and then extrapolating it to large pores. The proof is explained in the following.

We reproduced the adsorption of methane in cylindrical and slit pores (sizes ranging from $3-50\;\unit{nm}$) using GCMC simulations at $\SI{298}{K}$ and pressures of $2-10\unit{MPa}$. The simulations were run for $\num{e6}$ MC equilibration steps and $\num{e8}$ production steps. The chemical potential of bulk methane was set according to CoolProp under the defined conditions~\cite{CoolPropSI}. Solid-fluid and fluid-fluid interactions were the same as those of the MD simulations, and the effective surface density was set to $\rho_{\rm s}=\SI{38.19}{nm^{-2}}$ for both cylindrical and slit pores~\cite{Siderius2011}.

The excess adsorbed amount is defined as $N_{\rm ex}=(\rho_{\rm f}-\rho_{\rm b})V_{\rm f}$, where $\rho_{\rm f}$ is the density of confined methane and $\rho_{\rm b}$ is the density of bulk methane. $V_{\rm f}$ is the fluid-accessible volume in the pore.

\begin{figure}[H]
    \centering
    \subfloat[]{\includegraphics[width=0.49\textwidth]{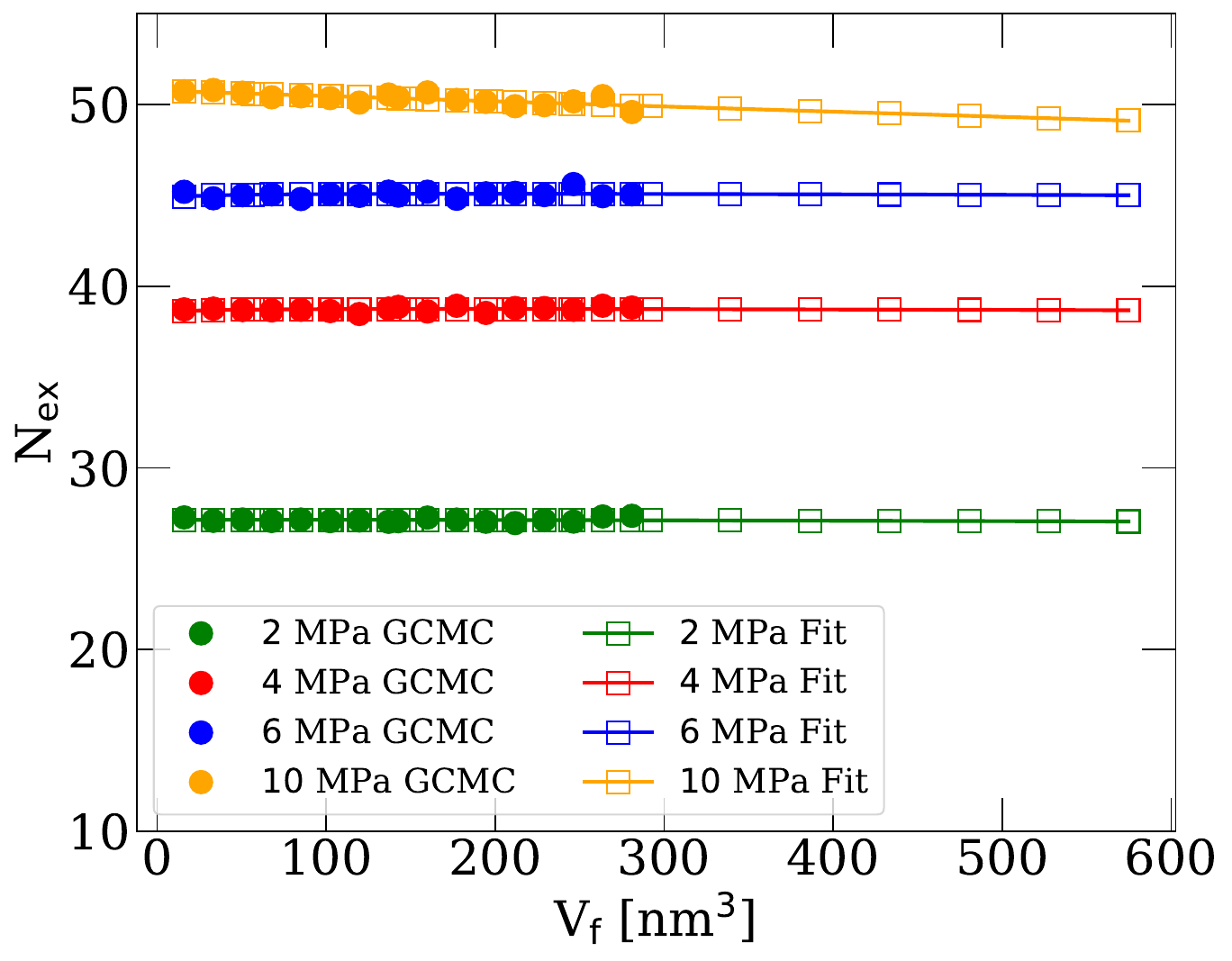}}
    \subfloat[]{\includegraphics[width=0.49\textwidth]{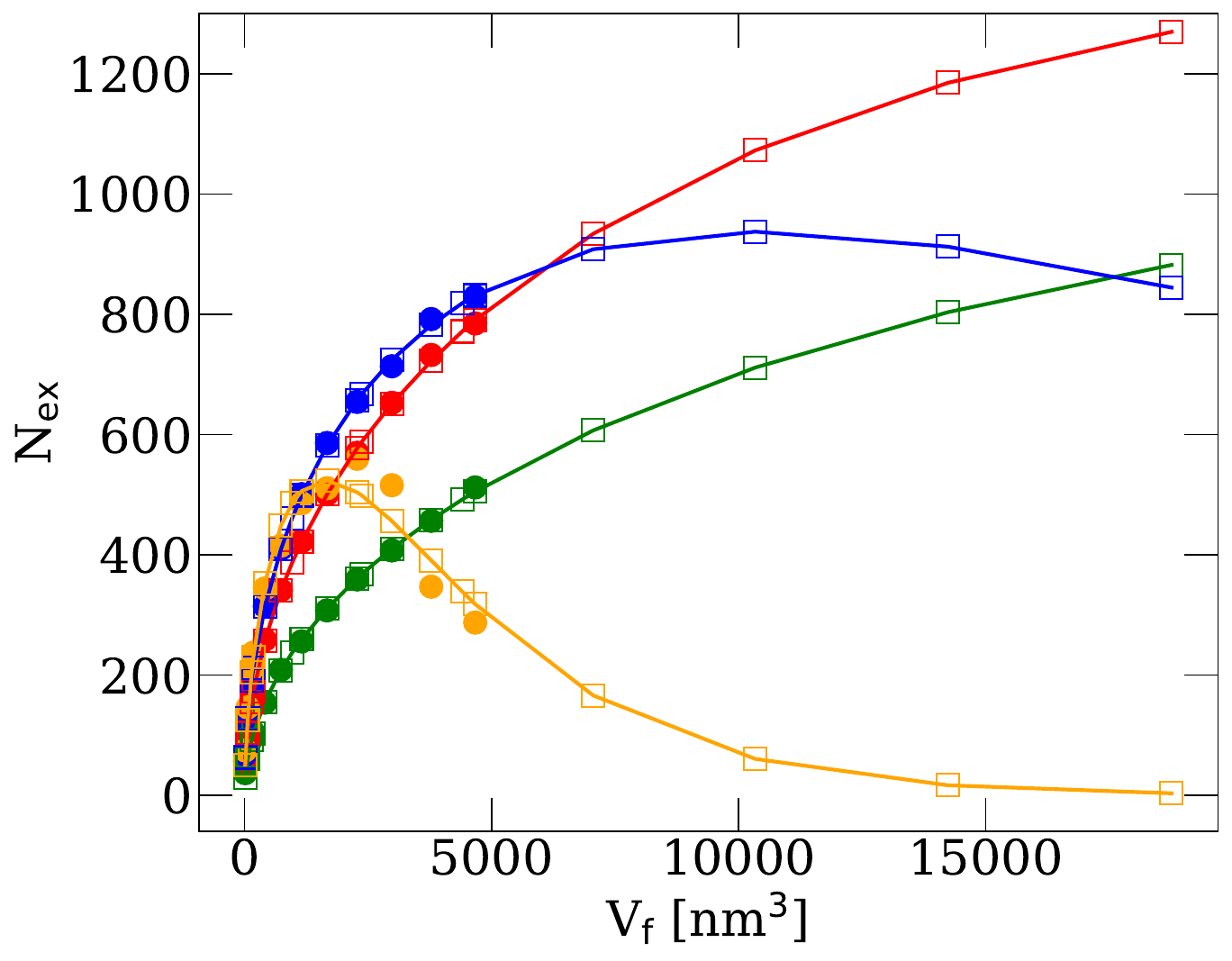}}
    \caption{\footnotesize{Predicted (filled circles) and estimated (empty squares, after fitting Eq. \ref{eq:SI_Nex}) excess adsorbed amount $N_{\rm ex}$ of methane in (a) slit pores and (b) cylindrical pores at pressures ranging from $2-10\unit{MPa}$ and $\SI{298}{K}$ as a function of fluid volume in the pore, $V_{\rm f}$. Simulated pores ranged from $3-50\;\unit{nm}$. Excess adsorbed amount was estimated from $\SI{3}{nm}$ to $\SI{100}{nm}$.}}
    \label{fig:SI_Nex}
\end{figure}

\begin{figure}[H]
    \centering
    \subfloat[]{\includegraphics[width=0.49\textwidth]{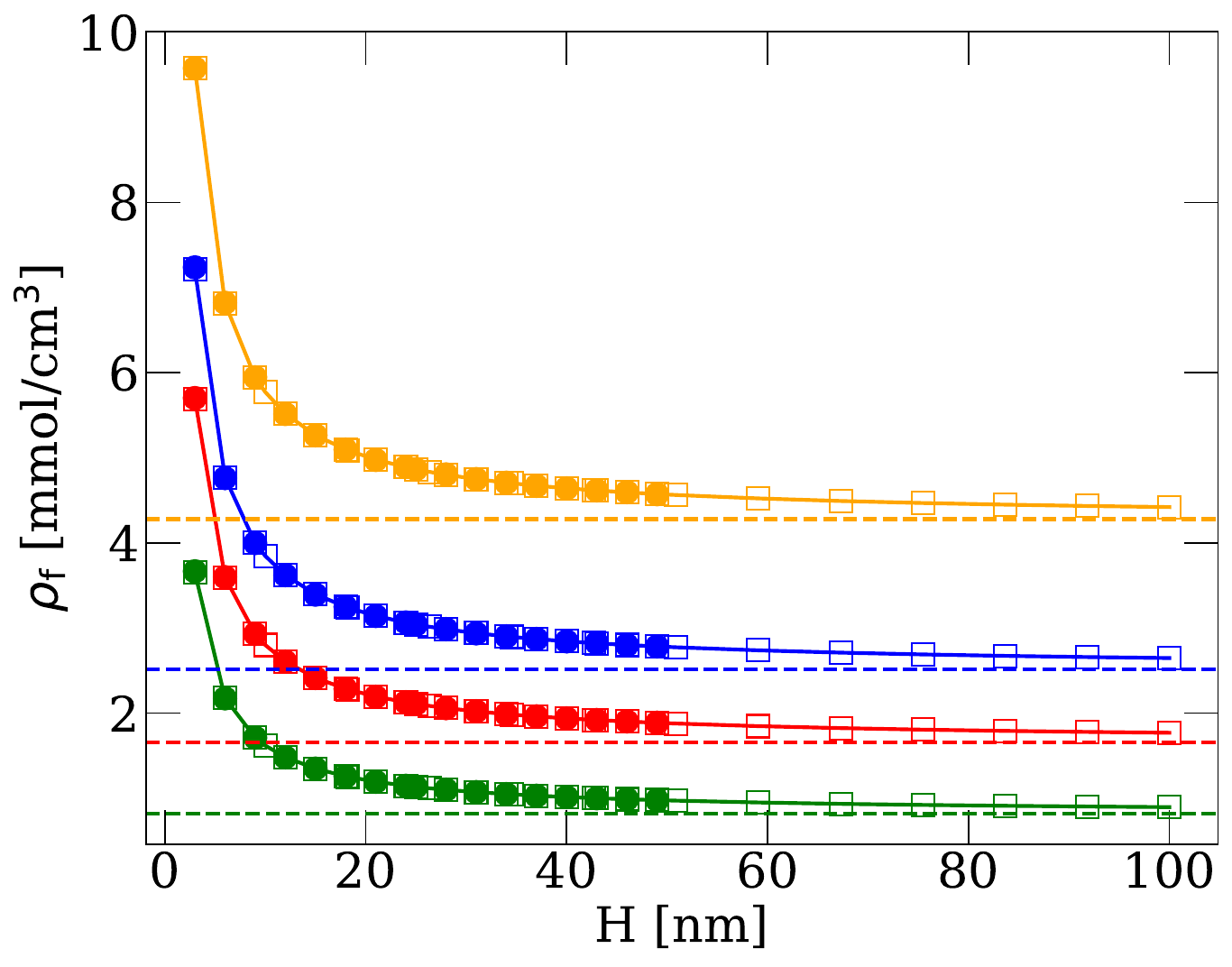}}
    \subfloat[]{\includegraphics[width=0.49\textwidth]{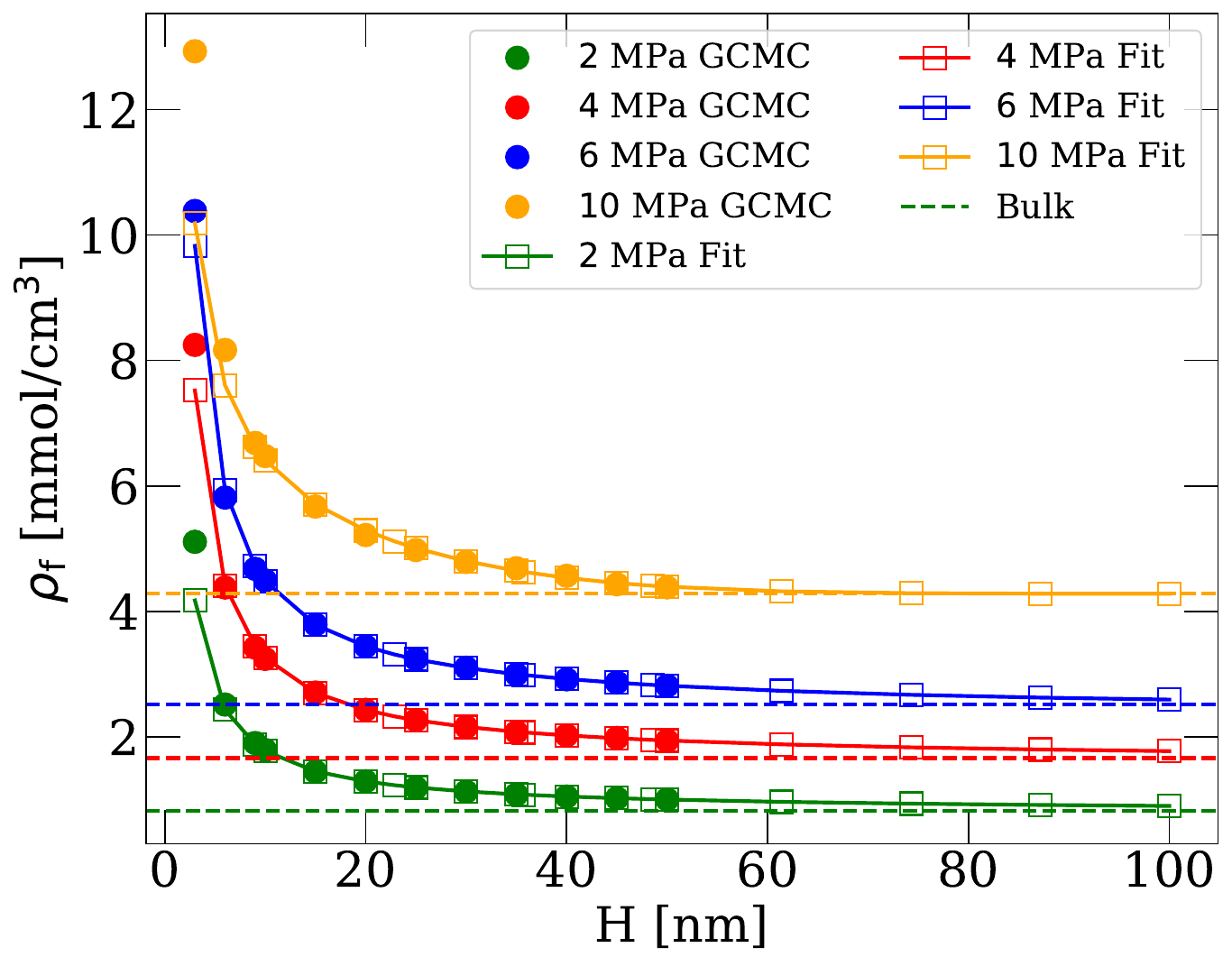}}
    \caption{\footnotesize{Predicted (filled circles) and estimated (empty squares, calculated from Eq. \ref{eq:SI_rhof_estimate}) fluid densities $\rho_{\rm f}$ of methane in (a) slit pores and (b) cylindrical pores at pressures ranging from $2-10\unit{MPa}$ and $\SI{298}{K}$ as a function of pore size. Simulated pores ranged from $3-50\;\unit{nm}$. Densities were estimated from $\SI{3}{nm}$ to $\SI{100}{nm}$. Dashed lines are the bulk densities of methane.}}
    \label{fig:SI_rhof}
\end{figure}

The predicted $N_{\rm ex}$ from GCMC simulations was then fitted according to:
\begin{equation}\label{eq:SI_Nex}
    N_{\rm ex}(V_{\rm f}; B, \alpha, \theta) = B\frac{V_{\rm f}^{\alpha-1}e^{-V_{\rm f}/\theta}}{\theta^{\alpha}\Gamma{(\alpha)}},
\end{equation}
where $B$, $\alpha$, and $\theta$ are the fitting parameters. The resulting density of the confined fluid is:
\begin{equation}\label{eq:SI_rhof_estimate}
    \rho_{\rm f} = \rho_{\rm b} + N_{\rm ex}/V_{\rm f}.
\end{equation}

Figure \ref{fig:SI_Nex} shows the predicted and excess adsorbed amount for slit and cylindrical pores. According to GCMC predictions, $N_{\rm ex}$ strongly depends on the pore shape. For slit pores, $N_{\rm ex}$ is mostly constant as a function of pore size. However, for cylindrical pores, $N_{\rm ex}$ reaches a maximum that depends on the thermodynamic conditions, i.e., the maximum shifts to lower pore sizes as the pressure increases. The Gamma distribution can be easily adjusted according to the pore shape by the shape parameter $\alpha$. The resulting values were $\alpha\approx1$ for slits and $\alpha\approx1.5$ for cylinders.

Once we have an estimate of the fluid density from Eq. \ref{eq:SI_rhof_estimate}, we can extrapolate it to a pore of $\SI{100}{nm}$ in size. Figure \ref{fig:SI_rhof} shows the predicted and estimated densities of methane in cylindrical and slit pores from $\SI{3}{nm}$ to $\SI{100}{nm}$. The estimated densities closely match predictions from GCMC simulations.

It is important to note that, for slit pores, Eq. \ref{eq:SI_Nex} succeeds in estimating $N_{\rm ex}$. However, for cylindrical pores, $N_{\rm ex}$ is underestimated at low pore sizes. However, as the pores become larger, $\rho_{\rm b}$ in Eq. \ref{eq:SI_rhof_estimate} becomes dominant in determining $\rho_{\rm f}$. Since our study focuses on large mesopores ($H \geq \SI{10}{nm}$), we can disregard the mismatches at $H < \SI{10}{nm}$.

\pagebreak
\subsection{Estimating Statistical Errors of Predicted Bulk Moduli} \label{sec:SI_StatErr}

We estimated the statistical error of $K_T$ through the bootstrap method~\cite{newman1999statistical}. This method takes $n$ random subsets of size $n$ (with replacement) of the collected data (the computed volume at every given timestep, or the computed number of molecules at every given MC step) to recompute the statistic $n$ times. Here, our statistic is the bulk modulus. If $n$ is large enough ($n > 100$), then the $n$ resamples will form a bootstrap distribution of the statistic with an estimated error:
\begin{equation}
    \varepsilon{[K_T]} = \sqrt{\Mean{K_T^2}-\Mean{K_T}^2}.
\end{equation}
Moreover, from the distribution, we can choose our confidence interval. In this work, we resampled $n = 500$ subsets from datasets of $\num{e4}$ values to estimate the error of $K_T$ with $\num{95}{\%}$ confidence.

\subsection{Fitting Parameters for Bulk Modulus Dependence on Inverse Pore Size} \label{sec:SI_fitKT_vs_invH}

The error of the fit of $K_T = K_T(H^{-1})$ was estimated as 

\begin{align}
\varepsilon{[K_T]}=\sqrt{(H^{-1})^{2b}(\varepsilon{[a]})^2 + ab(H^{-1})^{2(b-1)}(\varepsilon{[b]})^2 + (\varepsilon{[c]})^2}, 
\end{align}

\noindent where $(\varepsilon{[a]})^2$, $(\varepsilon{[b]})^2$, and $(\varepsilon{[c]})^2$ are the corresponding variances of the fitting parameters $a$, $b$ and $c$.

\begin{table*}[h]
\caption{\footnotesize{Results from fitting the bulk modulus vs pore size to the function $K_{T}(H) = aH^{-b} + c$ using data from cylindrical pores. Note that the parameter $c$ represents the bulk modulus of the non-confined fluid, $K_{\rm Bulk}$. $K_{\rm Bulk}$ was not included in the fitting data. R$^2$ is the coefficient of determination. The errors of the fitting parameters are their variances (errors lower than $\num{e-3}$ units were excluded).}}
\label{tab:SI_Cylinder_Fit_Parameters}
\begin{ruledtabular}
\begin{tabular}{lccccc}
$P_{\rm reservoir}$ (MPa) & a (\SI{}{nm \cdot MPa}) & b (unitless) & c (MPa) & $K_{\rm Bulk}$ (MPa) & $R^2$\\
\colrule
2 & 43.16 $\pm$ 1.35 & 1.17 $\pm$ 0.02 & 2.07 $\pm$ 0.06 & 1.92 $\pm$ 0.32 & 0.9982 \\
4 & 94.37 $\pm$ 2.43 & 1.24 $\pm$ 0.02 & 4.06 $\pm$ 0.09 & 3.65 $\pm$ 0.44 & 0.9987 \\
6 & 148.92 $\pm$ 4.90 & 1.32 $\pm$ 0.03 & 5.95 $\pm$ 0.16 & 5.35 $\pm$ 0.54 & 0.9983 \\
10 & 259.30 $\pm$ 4.17 & 1.31 $\pm$ 0.01 & 8.89 $\pm$ 0.12 & 8.50 $\pm$ 0.67 & 0.9998 \\
\end{tabular}
\end{ruledtabular}
\end{table*}

\begin{table*}[h]
\caption{\footnotesize{Results from fitting the bulk modulus vs pore size to the function $K_{T}(H) = aH^{-b} + c$ using data from slit pores. Note that the parameter $c$ represents the bulk modulus of the non-confined fluid, $K_{\rm Bulk}$. $K_{\rm Bulk}$ was not included in the fitting data. R$^2$ is the coefficient of determination. The errors of the fitting parameters are their variances (errors lower than $\num{e-3}$ units were excluded).}}
\label{tab:SI_Slit_Fit_Parameters}
\begin{ruledtabular}
\begin{tabular}{lccccc}
$P_{\rm reservoir}$ (MPa) & a (\SI{}{nm \cdot MPa}) & b (unitless) & c (MPa) & $K_{\rm Bulk}$ (MPa) & $R^2$\\
\colrule
2 & 27.13 $\pm$ 0.18 & 1.01 & 1.49 & 1.92 $\pm$ 0.32 & 0.9998\\
4 & 54.52 $\pm$ 3.2 & 1.12 & 3.32 $\pm$ 0.02 & 3.65 $\pm$ 0.44 & 0.9992 \\
6 & 75.42 $\pm$ 5.63 & 1.17 & 4.89 & 5.35 $\pm$ 0.54 & 0.9991\\
10 & 98.52 $\pm$ 4.04 & 1.20 & 8.14 $\pm$ 0.02& 8.50 $\pm$ 0.67 & 0.9997\\
\end{tabular}
\end{ruledtabular}
\end{table*}

\subsection{Finite-Size Dependence of the Bulk Modulus from Volume Fluctuations} \label{sec:SI_FiniteSizeKT}
\begin{figure}[H]
    \centering
    \subfloat[]{\includegraphics[width=0.49\textwidth]{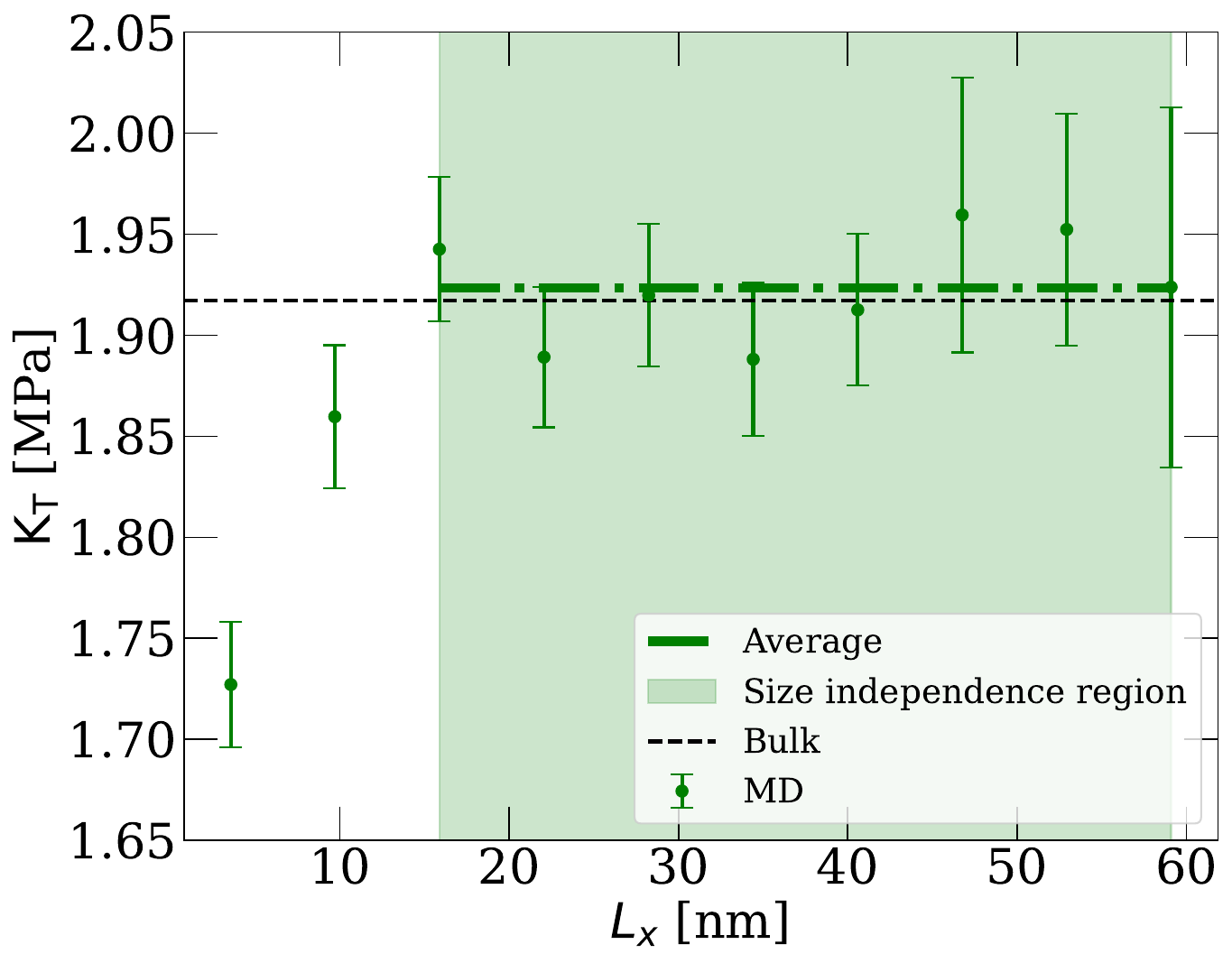}}
    \caption{\footnotesize{Predicted bulk modulus of methane in a \SI{59}{nm} slit pore at \SI{2}{MPa} and \SI{298}{K} as a function of the wall length, $L_{\rm x}$. The dashed line is the modulus of the bulk phase at the same thermodynamic conditions. The thick dashed-dotted line is the average of $K_T$ in the size-independent region ($L_{\rm x} \gtrsim \SI{16}{nm}$).}}
    \label{fig:SI_KT_vs_Lx}
\end{figure}

Finite-size effects may arise due to the dependence of $K_T$ on volume fluctuations. Hence, it is essential to analyze the dependence of $K_T$ on the box size along the periodic directions. Fig. \ref{fig:SI_KT_vs_Lx} shows the resulting $K_T$ of methane in a \SI{59}{nm} slit pore as a function of the wall length in the range $3r_{\rm cut} \le L_{\rm x} \le H$, where $r_{\rm cut} = \SI{1.2}{nm}$. Note that $L_{\rm x}$ comes from $A = L_{\rm x} \times L_{\rm x}$, where $A$ is the surface area of a wall. For a pore size of $H = \SI{59}{nm}$, it is expected to have $K_T$ approaching the modulus of its bulk phase, $K_{\rm Bulk}$. However, for sufficiently small wall lengths (in this case, $L_{\rm x} \lesssim \SI{16}{nm}$), the predicted $K_T$ underestimates $K_{\rm Bulk}$, while for large enough wall lengths ($L_{\rm x} \gtrsim \SI{16}{nm}$), our predictions fall around $K_{\rm Bulk}$. To predict fluid thermal properties, we must widely cover the phase space, so we can obtain a good estimate of the variance ($\Mean{\delta V}^2$ for our method). If our system is small in the amount of molecules or highly dense, we may not be able to sufficiently sample the phase space, resulting in underestimated predictions. For $L_{\rm x} = \SI{16}{nm}$, we simulated $\sim\num{8560}$ methane molecules. Although this effect has been observed mostly on MD predictions of transport properties, the dependence of our method on barostat protocols makes $K_T$ potentially sensitive to finite-size effects~\cite{Kim2018}.

Finite-size effects may also depend on the pore geometry. Unlike slit pores, we did not observe any such dependence in cylindrical pores for wall lengths $L_{\rm x} \ge 3r_{\rm cut}$.

\pagebreak


\begin{thebibliography}{28}%
\makeatletter
\providecommand \@ifxundefined [1]{%
 \@ifx{#1\undefined}
}%
\providecommand \@ifnum [1]{%
 \ifnum #1\expandafter \@firstoftwo
 \else \expandafter \@secondoftwo
 \fi
}%
\providecommand \@ifx [1]{%
 \ifx #1\expandafter \@firstoftwo
 \else \expandafter \@secondoftwo
 \fi
}%
\providecommand \natexlab [1]{#1}%
\providecommand \enquote  [1]{``#1''}%
\providecommand \bibnamefont  [1]{#1}%
\providecommand \bibfnamefont [1]{#1}%
\providecommand \citenamefont [1]{#1}%
\providecommand \href@noop [0]{\@secondoftwo}%
\providecommand \href [0]{\begingroup \@sanitize@url \@href}%
\providecommand \@href[1]{\@@startlink{#1}\@@href}%
\providecommand \@@href[1]{\endgroup#1\@@endlink}%
\providecommand \@sanitize@url [0]{\catcode `\\12\catcode `\$12\catcode
  `\&12\catcode `\#12\catcode `\^12\catcode `\_12\catcode `\%12\relax}%
\providecommand \@@startlink[1]{}%
\providecommand \@@endlink[0]{}%
\providecommand \url  [0]{\begingroup\@sanitize@url \@url }%
\providecommand \@url [1]{\endgroup\@href {#1}{\urlprefix }}%
\providecommand \urlprefix  [0]{URL }%
\providecommand \Eprint [0]{\href }%
\providecommand \doibase [0]{https://doi.org/}%
\providecommand \selectlanguage [0]{\@gobble}%
\providecommand \bibinfo  [0]{\@secondoftwo}%
\providecommand \bibfield  [0]{\@secondoftwo}%
\providecommand \translation [1]{[#1]}%
\providecommand \BibitemOpen [0]{}%
\providecommand \bibitemStop [0]{}%
\providecommand \bibitemNoStop [0]{.\EOS\space}%
\providecommand \EOS [0]{\spacefactor3000\relax}%
\providecommand \BibitemShut  [1]{\csname bibitem#1\endcsname}%
\let\auto@bib@innerbib\@empty
%</preamble>
\bibitem [{\citenamefont {Huber}(2015)}]{Huber2015}%
  \BibitemOpen
  \bibfield  {author} {\bibinfo {author} {\bibfnamefont {P.}~\bibnamefont
  {Huber}},\ }\bibfield  {title} {\bibinfo {title} {Soft matter in hard
  confinement: phase transition thermodynamics, structure, texture, diffusion
  and flow in nanoporous media (topical review)},\ }\href@noop {} {\bibfield
  {journal} {\bibinfo  {journal} {J. Phys.: Condens. Matter}\ }\textbf
  {\bibinfo {volume} {27}},\ \bibinfo {pages} {103102} (\bibinfo {year}
  {2015})}\BibitemShut {NoStop}%
\bibitem [{\citenamefont {Gubbins}\ \emph {et~al.}(2014)\citenamefont
  {Gubbins}, \citenamefont {Long},\ and\ \citenamefont
  {{\'S}liwinska-Bartkowiak}}]{Gubbins2014}%
  \BibitemOpen
  \bibfield  {author} {\bibinfo {author} {\bibfnamefont {K.~E.}\ \bibnamefont
  {Gubbins}}, \bibinfo {author} {\bibfnamefont {Y.}~\bibnamefont {Long}},\ and\
  \bibinfo {author} {\bibfnamefont {M.}~\bibnamefont
  {{\'S}liwinska-Bartkowiak}},\ }\bibfield  {title} {\bibinfo {title}
  {Thermodynamics of confined nano-phases},\ }\href@noop {} {\bibfield
  {journal} {\bibinfo  {journal} {J. Chem. Thermodyn.}\ }\textbf {\bibinfo
  {volume} {74}},\ \bibinfo {pages} {169} (\bibinfo {year} {2014})}\BibitemShut
  {NoStop}%
\bibitem [{\citenamefont {An}\ \emph {et~al.}(2023)\citenamefont {An},
  \citenamefont {Addington}, \citenamefont {Long}, \citenamefont {Rotnicki},
  \citenamefont {Śliwinska Bartkowiak}, \citenamefont {Thommes},\ and\
  \citenamefont {Gubbins}}]{An2023}%
  \BibitemOpen
  \bibfield  {author} {\bibinfo {author} {\bibfnamefont {R.}~\bibnamefont
  {An}}, \bibinfo {author} {\bibfnamefont {C.~K.}\ \bibnamefont {Addington}},
  \bibinfo {author} {\bibfnamefont {Y.}~\bibnamefont {Long}}, \bibinfo {author}
  {\bibfnamefont {K.}~\bibnamefont {Rotnicki}}, \bibinfo {author}
  {\bibfnamefont {M.}~\bibnamefont {Śliwinska Bartkowiak}}, \bibinfo {author}
  {\bibfnamefont {M.}~\bibnamefont {Thommes}},\ and\ \bibinfo {author}
  {\bibfnamefont {K.~E.}\ \bibnamefont {Gubbins}},\ }\bibfield  {title}
  {\bibinfo {title} {The nanoscale wetting parameter and its role in
  interfacial phenomena: Phase transitions in nanopores},\ }\href@noop {}
  {\bibfield  {journal} {\bibinfo  {journal} {Langmuir}\ }\textbf {\bibinfo
  {volume} {39}},\ \bibinfo {pages} {18730} (\bibinfo {year}
  {2023})}\BibitemShut {NoStop}%
\bibitem [{\citenamefont {Dobrzanski}\ \emph {et~al.}(2021)\citenamefont
  {Dobrzanski}, \citenamefont {Gurevich},\ and\ \citenamefont
  {Gor}}]{Dobrzanski2021}%
  \BibitemOpen
  \bibfield  {author} {\bibinfo {author} {\bibfnamefont {C.~D.}\ \bibnamefont
  {Dobrzanski}}, \bibinfo {author} {\bibfnamefont {B.}~\bibnamefont
  {Gurevich}},\ and\ \bibinfo {author} {\bibfnamefont {G.~Y.}\ \bibnamefont
  {Gor}},\ }\bibfield  {title} {\bibinfo {title} {Elastic properties of
  confined fluids from molecular modeling to ultrasonic experiments on porous
  solids},\ }\href@noop {} {\bibfield  {journal} {\bibinfo  {journal} {Appl.
  Phys. Rev.}\ }\textbf {\bibinfo {volume} {8}},\ \bibinfo {pages} {021317}
  (\bibinfo {year} {2021})}\BibitemShut {NoStop}%
\bibitem [{\citenamefont {Cheng}(2016)}]{cheng2016poroelasticity}%
  \BibitemOpen
  \bibfield  {author} {\bibinfo {author} {\bibfnamefont {A.~H.-D.}\
  \bibnamefont {Cheng}},\ }\href@noop {} {\emph {\bibinfo {title}
  {Poroelasticity}}},\ Vol.~\bibinfo {volume} {27}\ (\bibinfo  {publisher}
  {Springer},\ \bibinfo {year} {2016})\BibitemShut {NoStop}%
\bibitem [{\citenamefont {Page}\ \emph {et~al.}(1995)\citenamefont {Page},
  \citenamefont {Liu}, \citenamefont {Abeles}, \citenamefont {Herbolzheimer},
  \citenamefont {Deckman},\ and\ \citenamefont {Weitz}}]{Page1995}%
  \BibitemOpen
  \bibfield  {author} {\bibinfo {author} {\bibfnamefont {J.~H.}\ \bibnamefont
  {Page}}, \bibinfo {author} {\bibfnamefont {J.}~\bibnamefont {Liu}}, \bibinfo
  {author} {\bibfnamefont {B.}~\bibnamefont {Abeles}}, \bibinfo {author}
  {\bibfnamefont {E.}~\bibnamefont {Herbolzheimer}}, \bibinfo {author}
  {\bibfnamefont {H.~W.}\ \bibnamefont {Deckman}},\ and\ \bibinfo {author}
  {\bibfnamefont {D.~A.}\ \bibnamefont {Weitz}},\ }\bibfield  {title} {\bibinfo
  {title} {{Adsorption and desorption of a wetting fluid in Vycor studied by
  acoustic and optical techniques}},\ }\href@noop {} {\bibfield  {journal}
  {\bibinfo  {journal} {Phys. Rev. E}\ }\textbf {\bibinfo {volume} {52}},\
  \bibinfo {pages} {2763} (\bibinfo {year} {1995})}\BibitemShut {NoStop}%
\bibitem [{\citenamefont {Schappert}\ and\ \citenamefont
  {Pelster}(2013{\natexlab{a}})}]{Schappert2013JoP}%
  \BibitemOpen
  \bibfield  {author} {\bibinfo {author} {\bibfnamefont {K.}~\bibnamefont
  {Schappert}}\ and\ \bibinfo {author} {\bibfnamefont {R.}~\bibnamefont
  {Pelster}},\ }\bibfield  {title} {\bibinfo {title} {Elastic properties of
  liquid and solid argon in nanopores},\ }\href@noop {} {\bibfield  {journal}
  {\bibinfo  {journal} {J. Phys.: Condens. Matter}\ }\textbf {\bibinfo {volume}
  {25}},\ \bibinfo {pages} {415302} (\bibinfo {year}
  {2013}{\natexlab{a}})}\BibitemShut {NoStop}%
\bibitem [{\citenamefont {Schappert}\ and\ \citenamefont
  {Pelster}(2013{\natexlab{b}})}]{Schappert2013N2}%
  \BibitemOpen
  \bibfield  {author} {\bibinfo {author} {\bibfnamefont {K.}~\bibnamefont
  {Schappert}}\ and\ \bibinfo {author} {\bibfnamefont {R.}~\bibnamefont
  {Pelster}},\ }\bibfield  {title} {\bibinfo {title} {Strongly enhanced elastic
  modulus of solid nitrogen in nanopores},\ }\href@noop {} {\bibfield
  {journal} {\bibinfo  {journal} {Phys. Rev. B}\ }\textbf {\bibinfo {volume}
  {88}},\ \bibinfo {pages} {245443} (\bibinfo {year}
  {2013}{\natexlab{b}})}\BibitemShut {NoStop}%
\bibitem [{\citenamefont {Schappert}\ and\ \citenamefont
  {Pelster}(2014)}]{Schappert2014}%
  \BibitemOpen
  \bibfield  {author} {\bibinfo {author} {\bibfnamefont {K.}~\bibnamefont
  {Schappert}}\ and\ \bibinfo {author} {\bibfnamefont {R.}~\bibnamefont
  {Pelster}},\ }\bibfield  {title} {\bibinfo {title} {{Influence of the Laplace
  pressure on the elasticity of argon in nanopores}},\ }\href@noop {}
  {\bibfield  {journal} {\bibinfo  {journal} {Europhys. Lett.}\ }\textbf
  {\bibinfo {volume} {105}},\ \bibinfo {pages} {56001} (\bibinfo {year}
  {2014})}\BibitemShut {NoStop}%
\bibitem [{\citenamefont {Schappert}(2014)}]{Schappert2014thesis}%
  \BibitemOpen
  \bibfield  {author} {\bibinfo {author} {\bibfnamefont {K.}~\bibnamefont
  {Schappert}},\ }\emph {\bibinfo {title} {Confinement effects in nanopores:
  elastic properties, phase transitions, and sorption-induced deformation}},\
  \href@noop {} {Ph.D. thesis},\ \bibinfo  {school} {Universit\"at des
  Saarlandes} (\bibinfo {year} {2014})\BibitemShut {NoStop}%
\bibitem [{\citenamefont {Ogbebor}\ \emph {et~al.}(2023)\citenamefont
  {Ogbebor}, \citenamefont {Valenza}, \citenamefont {Ravikovitch},
  \citenamefont {Karunarathne}, \citenamefont {Muraro}, \citenamefont
  {Lebedev}, \citenamefont {Gurevich}, \citenamefont {Khalizov},\ and\
  \citenamefont {Gor}}]{Ogbebor2023}%
  \BibitemOpen
  \bibfield  {author} {\bibinfo {author} {\bibfnamefont {J.}~\bibnamefont
  {Ogbebor}}, \bibinfo {author} {\bibfnamefont {J.~J.}\ \bibnamefont
  {Valenza}}, \bibinfo {author} {\bibfnamefont {P.~I.}\ \bibnamefont
  {Ravikovitch}}, \bibinfo {author} {\bibfnamefont {A.}~\bibnamefont
  {Karunarathne}}, \bibinfo {author} {\bibfnamefont {G.}~\bibnamefont
  {Muraro}}, \bibinfo {author} {\bibfnamefont {M.}~\bibnamefont {Lebedev}},
  \bibinfo {author} {\bibfnamefont {B.}~\bibnamefont {Gurevich}}, \bibinfo
  {author} {\bibfnamefont {A.~F.}\ \bibnamefont {Khalizov}},\ and\ \bibinfo
  {author} {\bibfnamefont {G.~Y.}\ \bibnamefont {Gor}},\ }\bibfield  {title}
  {\bibinfo {title} {Ultrasonic study of water adsorbed in nanoporous
  glasses},\ }\href {https://doi.org/10.1103/PhysRevE.108.024802} {\bibfield
  {journal} {\bibinfo  {journal} {Phys. Rev. E}\ }\textbf {\bibinfo {volume}
  {108}},\ \bibinfo {pages} {024802} (\bibinfo {year} {2023})}\BibitemShut
  {NoStop}%
\bibitem [{\citenamefont {Schappert}\ and\ \citenamefont
  {Pelster}(2024)}]{Schappert2024}%
  \BibitemOpen
  \bibfield  {author} {\bibinfo {author} {\bibfnamefont {K.}~\bibnamefont
  {Schappert}}\ and\ \bibinfo {author} {\bibfnamefont {R.}~\bibnamefont
  {Pelster}},\ }\bibfield  {title} {\bibinfo {title} {Evaluating the pressure
  dependence of the longitudinal modulus of an adsorbate in nanopores using
  ultrasound: A novel procedure taking the effect of emptying pore ends into
  account},\ }\href@noop {} {\bibfield  {journal} {\bibinfo  {journal} {The
  Journal of Physical Chemistry C}\ }\textbf {\bibinfo {volume} {128}},\
  \bibinfo {pages} {21081} (\bibinfo {year} {2024})}\BibitemShut {NoStop}%
\bibitem [{\citenamefont {Didier}\ \emph {et~al.}(2025)\citenamefont {Didier},
  \citenamefont {Sam}, \citenamefont {Venegas},\ and\ \citenamefont
  {Coasne}}]{didier2025acoustic}%
  \BibitemOpen
  \bibfield  {author} {\bibinfo {author} {\bibfnamefont {L.}~\bibnamefont
  {Didier}}, \bibinfo {author} {\bibfnamefont {A.}~\bibnamefont {Sam}},
  \bibinfo {author} {\bibfnamefont {R.}~\bibnamefont {Venegas}},\ and\ \bibinfo
  {author} {\bibfnamefont {B.}~\bibnamefont {Coasne}},\ }\bibfield  {title}
  {\bibinfo {title} {Acoustic response of molecular adsorption and sound
  propagation in nanoporous materials},\ }\href@noop {} {\bibfield  {journal}
  {\bibinfo  {journal} {Physical Review Materials}\ }\textbf {\bibinfo {volume}
  {9}},\ \bibinfo {pages} {056001} (\bibinfo {year} {2025})}\BibitemShut
  {NoStop}%
\bibitem [{\citenamefont {Dobrzanski}\ \emph {et~al.}(2018)\citenamefont
  {Dobrzanski}, \citenamefont {Maximov},\ and\ \citenamefont
  {Gor}}]{Dobrzanski2018}%
  \BibitemOpen
  \bibfield  {author} {\bibinfo {author} {\bibfnamefont {C.~D.}\ \bibnamefont
  {Dobrzanski}}, \bibinfo {author} {\bibfnamefont {M.~A.}\ \bibnamefont
  {Maximov}},\ and\ \bibinfo {author} {\bibfnamefont {G.~Y.}\ \bibnamefont
  {Gor}},\ }\bibfield  {title} {\bibinfo {title} {Effect of pore geometry on
  the compressibility of a confined simple fluid},\ }\href@noop {} {\bibfield
  {journal} {\bibinfo  {journal} {J. Chem. Phys.}\ }\textbf {\bibinfo {volume}
  {148}},\ \bibinfo {pages} {054503} (\bibinfo {year} {2018})}\BibitemShut
  {NoStop}%
\bibitem [{\citenamefont {Maximov}\ and\ \citenamefont
  {Gor}(2018)}]{Maximov2018}%
  \BibitemOpen
  \bibfield  {author} {\bibinfo {author} {\bibfnamefont {M.~A.}\ \bibnamefont
  {Maximov}}\ and\ \bibinfo {author} {\bibfnamefont {G.~Y.}\ \bibnamefont
  {Gor}},\ }\bibfield  {title} {\bibinfo {title} {Molecular simulations shed
  light on potential uses of ultrasound in nitrogen adsorption experiments},\
  }\href@noop {} {\bibfield  {journal} {\bibinfo  {journal} {Langmuir}\
  }\textbf {\bibinfo {volume} {34}},\ \bibinfo {pages} {15650} (\bibinfo {year}
  {2018})}\BibitemShut {NoStop}%
\bibitem [{\citenamefont {Corrente}\ \emph {et~al.}(2020)\citenamefont
  {Corrente}, \citenamefont {Dobrzanski},\ and\ \citenamefont
  {Gor}}]{Corrente2020}%
  \BibitemOpen
  \bibfield  {author} {\bibinfo {author} {\bibfnamefont {N.~J.}\ \bibnamefont
  {Corrente}}, \bibinfo {author} {\bibfnamefont {C.~D.}\ \bibnamefont
  {Dobrzanski}},\ and\ \bibinfo {author} {\bibfnamefont {G.~Y.}\ \bibnamefont
  {Gor}},\ }\bibfield  {title} {\bibinfo {title} {Compressibility of
  supercritical methane in nanopores: A molecular simulation study},\
  }\href@noop {} {\bibfield  {journal} {\bibinfo  {journal} {Energy \& Fuels}\
  }\textbf {\bibinfo {volume} {34}},\ \bibinfo {pages} {1506} (\bibinfo {year}
  {2020})}\BibitemShut {NoStop}%
\bibitem [{\citenamefont {An}\ \emph {et~al.}(2025)\citenamefont {An},
  \citenamefont {Le}, \citenamefont {Gubbins}, \citenamefont
  {{\'S}liwinska-Bartkowiak},\ and\ \citenamefont {Thommes}}]{An2025}%
  \BibitemOpen
  \bibfield  {author} {\bibinfo {author} {\bibfnamefont {R.}~\bibnamefont
  {An}}, \bibinfo {author} {\bibfnamefont {R.}~\bibnamefont {Le}}, \bibinfo
  {author} {\bibfnamefont {K.~E.}\ \bibnamefont {Gubbins}}, \bibinfo {author}
  {\bibfnamefont {M.}~\bibnamefont {{\'S}liwinska-Bartkowiak}},\ and\ \bibinfo
  {author} {\bibfnamefont {M.}~\bibnamefont {Thommes}},\ }\bibfield  {title}
  {\bibinfo {title} {A perspective on characterization of porous materials: the
  melting line and triple point},\ }\href@noop {} {\bibfield  {journal}
  {\bibinfo  {journal} {Adsorption}\ }\textbf {\bibinfo {volume} {31}},\
  \bibinfo {pages} {69} (\bibinfo {year} {2025})}\BibitemShut {NoStop}%
\bibitem [{\citenamefont {Allen}\ and\ \citenamefont
  {Tildesley}(2017)}]{Allen2017}%
  \BibitemOpen
  \bibfield  {author} {\bibinfo {author} {\bibfnamefont {M.~P.}\ \bibnamefont
  {Allen}}\ and\ \bibinfo {author} {\bibfnamefont {D.~J.}\ \bibnamefont
  {Tildesley}},\ }\href@noop {} {\emph {\bibinfo {title} {Computer simulation
  of liquids}}}\ (\bibinfo  {publisher} {New York: Oxford},\ \bibinfo {year}
  {2017})\BibitemShut {NoStop}%
\bibitem [{\citenamefont {Siderius}\ and\ \citenamefont
  {Gelb}(2011)}]{Siderius2011}%
  \BibitemOpen
  \bibfield  {author} {\bibinfo {author} {\bibfnamefont {D.~W.}\ \bibnamefont
  {Siderius}}\ and\ \bibinfo {author} {\bibfnamefont {L.~D.}\ \bibnamefont
  {Gelb}},\ }\bibfield  {title} {\bibinfo {title} {{Extension of the Steele
  10-4-3 potential for adsorption calculations in cylindrical, spherical, and
  other pore geometries}},\ }\href@noop {} {\bibfield  {journal} {\bibinfo
  {journal} {J. Chem. Phys.}\ }\textbf {\bibinfo {volume} {135}},\ \bibinfo
  {pages} {084703} (\bibinfo {year} {2011})}\BibitemShut {NoStop}%
\bibitem [{\citenamefont {Maximov}\ \emph {et~al.}(2021)\citenamefont
  {Maximov}, \citenamefont {Molina},\ and\ \citenamefont {Gor}}]{Maximov2021}%
  \BibitemOpen
  \bibfield  {author} {\bibinfo {author} {\bibfnamefont {M.~A.}\ \bibnamefont
  {Maximov}}, \bibinfo {author} {\bibfnamefont {M.}~\bibnamefont {Molina}},\
  and\ \bibinfo {author} {\bibfnamefont {G.~Y.}\ \bibnamefont {Gor}},\
  }\bibfield  {title} {\bibinfo {title} {{The effect of interconnections on gas
  adsorption in materials with spherical mesopores: A Monte Carlo simulation
  study}},\ }\href@noop {} {\bibfield  {journal} {\bibinfo  {journal} {J. Chem.
  Phys.}\ }\textbf {\bibinfo {volume} {154}},\ \bibinfo {pages} {114706}
  (\bibinfo {year} {2021})}\BibitemShut {NoStop}%
\bibitem [{\citenamefont {Landers}\ \emph {et~al.}(2013)\citenamefont
  {Landers}, \citenamefont {Gor},\ and\ \citenamefont {Neimark}}]{Landers2013}%
  \BibitemOpen
  \bibfield  {author} {\bibinfo {author} {\bibfnamefont {J.}~\bibnamefont
  {Landers}}, \bibinfo {author} {\bibfnamefont {G.~Y.}\ \bibnamefont {Gor}},\
  and\ \bibinfo {author} {\bibfnamefont {A.~V.}\ \bibnamefont {Neimark}},\
  }\bibfield  {title} {\bibinfo {title} {Density functional theory methods for
  characterization of porous materials},\ }\href@noop {} {\bibfield  {journal}
  {\bibinfo  {journal} {Colloids Surf., A}\ }\textbf {\bibinfo {volume}
  {437}},\ \bibinfo {pages} {3} (\bibinfo {year} {2013})}\BibitemShut {NoStop}%
\bibitem [{\citenamefont {Tjatjopoulos}\ \emph {et~al.}(1988)\citenamefont
  {Tjatjopoulos}, \citenamefont {Feke},\ and\ \citenamefont
  {Mann~Jr}}]{Tjatjopoulos1988}%
  \BibitemOpen
  \bibfield  {author} {\bibinfo {author} {\bibfnamefont {G.~J.}\ \bibnamefont
  {Tjatjopoulos}}, \bibinfo {author} {\bibfnamefont {D.~L.}\ \bibnamefont
  {Feke}},\ and\ \bibinfo {author} {\bibfnamefont {J.~A.}\ \bibnamefont
  {Mann~Jr}},\ }\bibfield  {title} {\bibinfo {title} {Molecule-micropore
  interaction potentials},\ }\href@noop {} {\bibfield  {journal} {\bibinfo
  {journal} {J. Phys. Chem.}\ }\textbf {\bibinfo {volume} {92}},\ \bibinfo
  {pages} {4006} (\bibinfo {year} {1988})}\BibitemShut {NoStop}%
\bibitem [{\citenamefont {Thompson}\ \emph {et~al.}(2022)\citenamefont
  {Thompson}, \citenamefont {Aktulga}, \citenamefont {Berger}, \citenamefont
  {Bolintineanu}, \citenamefont {Brown}, \citenamefont {Crozier}, \citenamefont
  {in~'t Veld}, \citenamefont {Kohlmeyer}, \citenamefont {Moore}, \citenamefont
  {Nguyen}, \citenamefont {Shan}, \citenamefont {Stevens}, \citenamefont
  {Tranchida}, \citenamefont {Trott},\ and\ \citenamefont {Plimpton}}]{LAMMPS}%
  \BibitemOpen
  \bibfield  {author} {\bibinfo {author} {\bibfnamefont {A.~P.}\ \bibnamefont
  {Thompson}}, \bibinfo {author} {\bibfnamefont {H.~M.}\ \bibnamefont
  {Aktulga}}, \bibinfo {author} {\bibfnamefont {R.}~\bibnamefont {Berger}},
  \bibinfo {author} {\bibfnamefont {D.~S.}\ \bibnamefont {Bolintineanu}},
  \bibinfo {author} {\bibfnamefont {W.~M.}\ \bibnamefont {Brown}}, \bibinfo
  {author} {\bibfnamefont {P.~S.}\ \bibnamefont {Crozier}}, \bibinfo {author}
  {\bibfnamefont {P.~J.}\ \bibnamefont {in~'t Veld}}, \bibinfo {author}
  {\bibfnamefont {A.}~\bibnamefont {Kohlmeyer}}, \bibinfo {author}
  {\bibfnamefont {S.~G.}\ \bibnamefont {Moore}}, \bibinfo {author}
  {\bibfnamefont {T.~D.}\ \bibnamefont {Nguyen}}, \bibinfo {author}
  {\bibfnamefont {R.}~\bibnamefont {Shan}}, \bibinfo {author} {\bibfnamefont
  {M.~J.}\ \bibnamefont {Stevens}}, \bibinfo {author} {\bibfnamefont
  {J.}~\bibnamefont {Tranchida}}, \bibinfo {author} {\bibfnamefont
  {C.}~\bibnamefont {Trott}},\ and\ \bibinfo {author} {\bibfnamefont {S.~J.}\
  \bibnamefont {Plimpton}},\ }\bibfield  {title} {\bibinfo {title} {{LAMMPS} -
  a flexible simulation tool for particle-based materials modeling at the
  atomic, meso, and continuum scales},\ }\href
  {https://doi.org/10.1016/j.cpc.2021.108171} {\bibfield  {journal} {\bibinfo
  {journal} {Comp. Phys. Comm.}\ }\textbf {\bibinfo {volume} {271}},\ \bibinfo
  {pages} {108171} (\bibinfo {year} {2022})}\BibitemShut {NoStop}%
\bibitem [{\citenamefont {Martin}\ and\ \citenamefont
  {Siepmann}(1998)}]{Martin1998}%
  \BibitemOpen
  \bibfield  {author} {\bibinfo {author} {\bibfnamefont {M.~G.}\ \bibnamefont
  {Martin}}\ and\ \bibinfo {author} {\bibfnamefont {J.~I.}\ \bibnamefont
  {Siepmann}},\ }\bibfield  {title} {\bibinfo {title} {{Transferable potentials
  for phase equilibria. 1. United-atom description of n-alkanes}},\ }\href@noop
  {} {\bibfield  {journal} {\bibinfo  {journal} {J. Phys. Chem. B}\ }\textbf
  {\bibinfo {volume} {102}},\ \bibinfo {pages} {2569} (\bibinfo {year}
  {1998})}\BibitemShut {NoStop}%
\bibitem [{\citenamefont {Mosher}\ \emph {et~al.}(2013)\citenamefont {Mosher},
  \citenamefont {He}, \citenamefont {Liu}, \citenamefont {Rupp},\ and\
  \citenamefont {Wilcox}}]{Mosher2013}%
  \BibitemOpen
  \bibfield  {author} {\bibinfo {author} {\bibfnamefont {K.}~\bibnamefont
  {Mosher}}, \bibinfo {author} {\bibfnamefont {J.}~\bibnamefont {He}}, \bibinfo
  {author} {\bibfnamefont {Y.}~\bibnamefont {Liu}}, \bibinfo {author}
  {\bibfnamefont {E.}~\bibnamefont {Rupp}},\ and\ \bibinfo {author}
  {\bibfnamefont {J.}~\bibnamefont {Wilcox}},\ }\bibfield  {title} {\bibinfo
  {title} {Molecular simulation of methane adsorption in micro-and mesoporous
  carbons with applications to coal and gas shale systems},\ }\href@noop {}
  {\bibfield  {journal} {\bibinfo  {journal} {Int. J. Coal Geol.}\ }\textbf
  {\bibinfo {volume} {109}},\ \bibinfo {pages} {36} (\bibinfo {year}
  {2013})}\BibitemShut {NoStop}%
\bibitem [{\citenamefont {Lorentz}(1881)}]{Lorentz1881}%
  \BibitemOpen
  \bibfield  {author} {\bibinfo {author} {\bibfnamefont {H.~A.}\ \bibnamefont
  {Lorentz}},\ }\bibfield  {title} {\bibinfo {title} {Ueber die anwendung des
  satzes vom virial in der kinetischen theorie der gase},\ }\href
  {https://doi.org/10.1002/andp.18812480110} {\bibfield  {journal} {\bibinfo
  {journal} {Annalen der Physik}\ }\textbf {\bibinfo {volume} {248}},\ \bibinfo
  {pages} {127–136} (\bibinfo {year} {1881})}\BibitemShut {NoStop}%
\bibitem [{\citenamefont {Sun}\ \emph {et~al.}(2019)\citenamefont {Sun},
  \citenamefont {Kang},\ and\ \citenamefont {Kang}}]{Sun2019density}%
  \BibitemOpen
  \bibfield  {author} {\bibinfo {author} {\bibfnamefont {Z.}~\bibnamefont
  {Sun}}, \bibinfo {author} {\bibfnamefont {Y.}~\bibnamefont {Kang}},\ and\
  \bibinfo {author} {\bibfnamefont {Y.}~\bibnamefont {Kang}},\ }\bibfield
  {title} {\bibinfo {title} {Density functional study on enhancement of modulus
  of confined fluid in nanopores},\ }\href@noop {} {\bibfield  {journal}
  {\bibinfo  {journal} {Ind. Eng. Chem. Res.}\ }\textbf {\bibinfo {volume}
  {58}},\ \bibinfo {pages} {15637} (\bibinfo {year} {2019})}\BibitemShut
  {NoStop}%
\bibitem [{\citenamefont {Boerner}\ \emph {et~al.}(2023)\citenamefont
  {Boerner}, \citenamefont {Deems}, \citenamefont {Furlani}, \citenamefont
  {Knuth},\ and\ \citenamefont {Towns}}]{Boerner2023}%
  \BibitemOpen
  \bibfield  {author} {\bibinfo {author} {\bibfnamefont {T.~J.}\ \bibnamefont
  {Boerner}}, \bibinfo {author} {\bibfnamefont {S.}~\bibnamefont {Deems}},
  \bibinfo {author} {\bibfnamefont {T.~R.}\ \bibnamefont {Furlani}}, \bibinfo
  {author} {\bibfnamefont {S.~L.}\ \bibnamefont {Knuth}},\ and\ \bibinfo
  {author} {\bibfnamefont {J.}~\bibnamefont {Towns}},\ }\bibfield  {title}
  {\bibinfo {title} {{{ACCESS}}: {{Advancing Innovation}}: {{NSF}}'s {{Advanced
  Cyberinfrastructure Coordination Ecosystem}}: {{Services}} \& {{Support}}},\
  }in\ \href {https://doi.org/10.1145/3569951.3597559} {\emph {\bibinfo
  {booktitle} {Practice and {{Experience}} in {{Advanced Research
  Computing}}}}},\ \bibinfo {series and number} {{{PEARC}} '23}\ (\bibinfo
  {publisher} {Association for Computing Machinery},\ \bibinfo {address} {New
  York, NY, USA},\ \bibinfo {year} {2023})\ pp.\ \bibinfo {pages}
  {173--176}\BibitemShut {NoStop}%
  \bibitem [{\citenamefont {Bell}\ \emph {et~al.}(2014)\citenamefont {Bell},
  \citenamefont {Wronski}, \citenamefont {Quoilin},\ and\ \citenamefont
  {Lemort}}]{CoolPropSI}%
  \BibitemOpen
  \bibfield  {author} {\bibinfo {author} {\bibfnamefont {I.~H.}\ \bibnamefont
  {Bell}}, \bibinfo {author} {\bibfnamefont {J.}~\bibnamefont {Wronski}},
  \bibinfo {author} {\bibfnamefont {S.}~\bibnamefont {Quoilin}},\ and\ \bibinfo
  {author} {\bibfnamefont {V.}~\bibnamefont {Lemort}},\ }\bibfield  {title}
  {\bibinfo {title} {{Pure and Pseudo-pure Fluid Thermophysical Property
  Evaluation and the Open-Source Thermophysical Property Library CoolProp}},\
  }\href {https://doi.org/10.1021/ie4033999} {\bibfield  {journal} {\bibinfo
  {journal} {Ind. Eng. Chem. Res.}\ }\textbf {\bibinfo {volume} {53}},\
  \bibinfo {pages} {2498} (\bibinfo {year} {2014})}\BibitemShut {NoStop}%
\bibitem [{\citenamefont {Newman}\ and\ \citenamefont
  {Barkema}(1999)}]{newman1999statistical}%
  \BibitemOpen
  \bibfield  {author} {\bibinfo {author} {\bibfnamefont {M.~E.~J.}\
  \bibnamefont {Newman}}\ and\ \bibinfo {author} {\bibfnamefont {G.~T.}\
  \bibnamefont {Barkema}},\ }\href@noop {} {\emph {\bibinfo {title} {Monte
  Carlo Methods in Statistical Physics}}}\ (\bibinfo  {publisher} {Oxford
  University Press},\ \bibinfo {year} {1999})\ Chap.\ \bibinfo {chapter}
  {3.4.3}\BibitemShut {NoStop}%
\bibitem [{\citenamefont {Kim}\ \emph {et~al.}(2018)\citenamefont {Kim},
  \citenamefont {Han}, \citenamefont {Kim}, \citenamefont {Li}, \citenamefont
  {Karniadakis},\ and\ \citenamefont {Lee}}]{Kim2018}%
  \BibitemOpen
  \bibfield  {author} {\bibinfo {author} {\bibfnamefont {K.-S.}\ \bibnamefont
  {Kim}}, \bibinfo {author} {\bibfnamefont {M.~H.}\ \bibnamefont {Han}},
  \bibinfo {author} {\bibfnamefont {C.}~\bibnamefont {Kim}}, \bibinfo {author}
  {\bibfnamefont {Z.}~\bibnamefont {Li}}, \bibinfo {author} {\bibfnamefont
  {G.~E.}\ \bibnamefont {Karniadakis}},\ and\ \bibinfo {author} {\bibfnamefont
  {E.~K.}\ \bibnamefont {Lee}},\ }\bibfield  {title} {\bibinfo {title} {Nature
  of intrinsic uncertainties in equilibrium molecular dynamics estimation of
  shear viscosity for simple and complex fluids},\ }\href
  {https://doi.org/10.1063/1.5035119} {\bibfield  {journal} {\bibinfo
  {journal} {J. Chem. Phys.}\ }\textbf {\bibinfo {volume} {149}},\ \bibinfo
  {pages} {044510} (\bibinfo {year} {2018})}\BibitemShut {NoStop}%
\end{thebibliography}
\end{document}